\documentclass{aastex6}

\newcommand{\degree}{\ensuremath{^{\circ}}}
\newcommand{\kms}{km~s\ensuremath{^{-1}}}
\newcommand{\vlsr}{\ensuremath{v_{\rm LSR}}}
\newcommand{\hi}{\ion{H}{1}}
\newcommand{\hii}{\ion{H}{2}}
\newcommand{\hei}{[\ion{He}{1}]}

 \newcommand{\oiii}{[\ion{O}{3}]}
\newcommand{\sii}{[\ion{S}{2}]}
\newcommand{\nii}{[\ion{N}{2}]}
\newcommand{\ha}{H\ensuremath{\alpha}}
\newcommand{\hb}{H\ensuremath{\beta}}
\newcommand{\iha}{\ensuremath{I_{\mathrm{H}\alpha}}}

\newcommand{\lo}{\ensuremath{\lambda}~Ori}
\newcommand{\lc}{\ensuremath{L_c}}

\shorttitle{Ionization Structure of Sharpless 2-264}
\shortauthors{Sahan \& Haffner}

\begin{document}

\title{The Ionization Structure of Sharpless 2-264: Multiwavelength Observations of the \lo\ \hii\ Region}


\author{M. Sahan}
\affil{Department of Physics, University of Osmaniye Korkut Ata, Osmaniye, 80000, Turkey }
\email{msahan2000@yahoo.com}

 \and
 
\author{L. M. Haffner}
\affil{Department of Astronomy, University of Wisconsin, 475 North Charter Street, Madison, WI 53706}
\affil{Space Science Institute, 4750 Walnut Street, Suite 205, Boulder, CO 80301}
\email{haffner@astro.wisc.edu}


\begin{abstract}

We present velocity-resolved maps taken with the Wisconsin H-Alpha Mapper (WHAM) in \ha, \sii\ $\lambda6716$, and \nii\ $\lambda6583$ around the well-known O8~III star $\lambda$ Ori A (HD 36861) ($\ell = 185\degree$ to $205\degree$, $b = -24\degree$ to $-1\degree$). The integrated intensity ($\vlsr = -80$ to $+80$ \kms), $\iha$, within WHAM's one-degree beams varies from $\sim 190$ R near the center to $\sim 10$ R on the periphery of the nebula where it becomes comparable to foreground and/or background emission in this complex region. Intensity ratios for \nii/\ha\ and \sii/\ha\ average 0.28 and 0.35, respectively. In both ratios, higher values are found preferentially at larger radii from \lo, although the behavior of \nii/\ha\ is complicated near the edges of the nebula. The \sii/\nii\ intensity ratio ranges from $\sim 0.5$ to $\sim 1.0$, with the value increasing toward larger radii (and lower \ha\ intensities). Variations of \sii/\ha, \nii/\ha, and \sii/\nii\ line ratios in this diffuse region show some similar trends to those seen in the warm ionized medium (WIM) but with generally lower metal-line ratios. As with the WIM, the trends are driven by changes in the underlying physical parameters, most notably the  ionization states and gas temperature. To investigate which cause might be dominant in this region, we use these extremely high signal-to-noise observations to construct a map of temperature and non-thermal velocity throughout the nebula. Using the line widths of \ha\ and \sii, we separate thermal and non-thermal components and find spatial trends of these parameters within the nebula. Ion temperatures range between 4000~K and 8000~K throughout the nebula. The non-thermal velocity map reveals a decrease in velocity from about 10 \kms\ to 5 \kms\ from the center to the edge of the lower half of the \hii\ region. In addition to using the widths as a measure of temperature, we also use the variation in \nii/\ha\ to estimate electron temperature. The results obtained from this diffuse \hii\ region around \lo\ can be compared to studies of the WIM to provide important insight into the nature of the diffuse ionized gas (DIG) throughout the disk and halo of the Galaxy.

\end{abstract}

\keywords{H II regions --- ISM: structure --- ISM: individual (Sh 2-264) --- ISM: atoms --- stars: individual (HD 36861)}

\section{INTRODUCTION}
\label{sec:intro}

The Wisconsin H-Alpha Mapper (WHAM) has measured a variety of
optical emission lines for a multitude of large-scale ionized
structures that include loops, filaments, and bubbles. The results
from such studies provide insight into the workings of these
discrete structures as well as the warm ionized medium (WIM) that
pervades the Galaxy.  The unique design of its 15-cm dual-etalon
Fabry-Perot spectrometer allows WHAM to produce the first kinematic
\ha\ map of the ionized gas in the Milky Way \citep{haf03} as well
as spectral maps of diagnostic lines such as \sii\ $\lambda$6716,
\nii\ $\lambda$6583, and \oiii\ $\lambda$5007
 \citep[e.g.,][]{haf99,mad06}. Other recent large-angle surveys that trace diffuse \ha\ in our region of study include the Southern \ha\ Sky Survey Atlas (SHASSA; \citealt{SHASSA}) and the Virginia Tech Spectral-Line Survey (VTSS; \citealt{VTSS}). Although both provide higher spatial resolution than WHAM, these imaging surveys provide no spectral information and cannot isolate overlapping emission features. WHAM is also an order of magnitude more sensitive in the survey-mode presented here due to its large beam (1\arcdeg) sampling and its ability to spectrally remove all contamination from atmospheric sky lines. Here, we present the
first multi-wavelength spectral maps of Sh 2-264, an \hii\ region
ionized by the O8~III star, \lo\ A (HD 36861; $\ell = 195\fdg05$, $b
= -12\fdg00$).

Such relatively isolated, large-angle, diffuse \hii\ regions form an
important observational bridge between the concentrated, classical,
bright nebulae in high-density star-forming environments and the
wide-spread, low-density WIM. Unlike the WIM, these regions have
simple, well-understood sources of ionization with density
characteristics more typical of the diffuse ISM. A few of these have
also become important calibration targets for WHAM as regions of
moderate intensity over large-angular scales that are visible from
its former northern and new southern locations.

\lo\ A anchors a small knot of massive stars in the center of the
\hii\ region but dominates the ionizing luminosity of the
association. Studies of the recent and on-going star formation in
the immediate surroundings of \lo\ suggest that violent stellar
activity has reorganized the gas since its formation. A
comprehensive review of the region, with particular emphasis on the
star-formation history, is presented by \citet{mat08}. For our
study, we adopt the hypothesis put forth in \citet{dol01,dol02} that
at least one supernova has reorganized the molecular gas in the
region 1--2 Myr ago, truncating star formation in the central region
near \lo\ and creating the low-density cavity for the diffuse \hii\
region. We also adopt the main-sequence fitting distance for the
cluster of $450 \pm 50$ pc as derived by  \citet{dol01}  as the
distance to the center of Sh~2-264, placing it about 94~pc away from
the Galactic Plane. Although the \hii\ region appears to be
relatively isolated and dominated predominantly by the influence
from \lo\ alone, Sh 2-264 is in the vicinity of the impressive
Orion-Eridanus superbubble complex and likely shares a complex,
common history with the region.  \citet{Ochs15}  present a recent,
comprehensive view derived from several multiwavelength surveys.

Using WHAM observations of the \ha\ line ($\lambda6563$) and the
faint optical emission lines \sii\ $\lambda$6716 and \nii\
$\lambda$6583, we begin a study of the physical state of
the diffuse \hii\ region surrounding \lo. \ha\ survey data are
combined with \sii\ and \nii\ observations to learn about the
ionization state of the gas and to derive non-thermal velocity and
temperature throughout the $\sim 62$-pc-diameter \hii\ region. Our
multiwavelength study also probes into regions beyond the apparent
Str\"{o}mgren sphere, revealing nearby filamentary structure,
including a portion of Barnard's Loop (Sh 2-276).

\section{OBSERVATIONS}
\label{sec:obs}

The WHAM spectrometer was designed to detect optical emission lines
from diffuse sources at high spectral resolution. The recombination
line \ha---typically the brightest of these---is used as the primary
tracer to map ionized gas. The physical conditions and ionization
state of the gas can then be studied by adding observations of
collisionally excited diagnostic lines. WHAM's first major effort
produced the first kinematic map of the northern sky ($\delta \ge
-30\degree$) in \ha, which took approximately two years. This map
reveals detailed structure of supernova remnants in the form of
giant bubbles and super-shells as well as filaments of warm
pervasive gas seen far from classical \hii\ regions. The WHAM
Northern Sky Survey \citep[WHAM-NSS;][]{haf03}, containing more than
37,500 individual \ha\ spectra obtained with a 1\degree\-diameter
beam, is the first spectral survey to map the spatial distribution
and kinematic structure of diffuse ionized gas throughout the Galaxy
\citep{rey02,haf03}. By using a Fabry-Perot spectrometer, WHAM can accept this large solid angle while delivering high spectral resolution ($R \approx 25,000$). This beam size combined with our ability to cleanly separate Galactic and atmospheric emission allows the deepest WHAM observations to be roughly 100 times more sensitive to diffuse, extended emission than current imaging instruments, opening up new opportunities for exploration. It is now located in Chile and has observed the southern sky to provide the first all-sky, kinematic \ha\ map analogous
to surveys of \hi\ in 21 cm. Considerable detail about WHAM's
optical design, performance, and data reduction can be found in
\citet{tuf97} and \citet{haf99}.

In this study, \ha\ observations have been extracted from the
WHAM-NSS and have been supplemented with WHAM observations in \sii\
and \nii\ within a $20\degree \times 24\degree$ area surrounding
\lo. The area surveyed spans the region in Galactic coordinates
$\ell = 185\degree$ to $\ell = 205\degree$ and $b = -24\degree$ to
$b = -1\degree$. The \ha\ observations were mainly taken in 1997
during the northern \ha\ survey while the \sii\ and \nii\
observations were obtained in 2006. On-sky exposure times are 30 s for
\ha\ and 60 s for \sii\ and \nii\ observations. Since the WHAM
spectrometer records only spectral information, each observation
taken with WHAM represents the average spectrum within a 1\degree-diameter beam on the sky. The region of study includes 552 original observations in each spectral line. Five directions contain too much spectral contamination from bright stars having $V < 3$ mag ($\alpha$, $\gamma$, $\epsilon$, and $\delta$ Ori; $\zeta$ Tau) and are excluded from the final dataset of 547 spectra in each line.

Due to heavy dilution by the one-degree beam relative to diffuse emission, only the brightest stars impact our observations. However, all spectra obtained with WHAM contain noticeable features from the earth's atmosphere. In \ha,  the geocoronal \ha\ emission line is the primary contaminant, which arises from neutral hydrogen being excited by
scattered solar Ly$\beta$ radiation in the exosphere
\citep{mie06,nos06}. \sii\ observations contain emission
from the terrestrial neon line (\ion{Ne}{1} $\lambda6717$), which
originates from city lights scattered into the night sky around Kitt
Peak \citep{rey82,haf99}.  This line is
located $+26.8$ \kms\ from the geocentric zero of \sii\ emission but is much fainter than geocoronal \ha.
Since these lines vary independently from the fainter atmospheric lines described below, a
Gaussian profile is fit to them and subtracted from each observation. \citet{haf03} describes this procedure for \ha\ in more depth.

\sii\ and \nii\ emission---as well as \ha---are also contaminated by $
\sim 5$--7 other atmospheric lines which have
full-width-at-half-maximum (FWHM) $ \sim 10$ \kms\ and intensities
$I \sim 0.05$--0.5 R. The absolute strengths of the atmospheric
lines are observed to vary with position and time. Within a given
night, these intensities correlate well with zenith distance. Yet
the relative strengths of these faint lines are constant to $ \sim
10\%$. As a result, we can construct an atmospheric template by
observing directions with little Galactic emission multiple times
over several nights \citep{hau02}. We use high signal-to-noise averaged
spectrum from each observational ``block'' (up to a $7 \times 7$
grid of individual 1\degree-observations) to determine the absolute
intensity level of the template for that block. Semi-automated
software is then used to fit Gaussian profiles to multiple Galactic
components and remove atmospheric emission present in each
individual spectrum. \citet{haf03} describe this procedure in more detail for the \ha\ survey. For all three spectral windows presented here, the \emph{total} emission from faint atmospheric lines in our 200 \kms\ spectral window near $\vlsr = 0$ \kms\ varies from about 0.2 to 0.8 R depending on the night and look direction. We are able to determine the absolute intensity level of the atmospheric templates better than 10\% in most spectra. Directions where the Galactic emission is very bright ($> 20$ R) make this determination difficult, but for the same reason any residual contamination from atmospheric lines also becomes negligible. From our analysis, we estimate a conservative systematic uncertainty of 0.05 R contributed by residuals over the whole 200 \kms\ spectrum. 

Each spectrum is calibrated to obtain a stable velocity zero-point with
respect to the local standard of rest (LSR). Velocity calibration in
\ha\ and \sii\ is accomplished by using the brighter atmospheric lines (geocoronal \ha\ and terrestrial Ne I). Line centers  fit during atmospheric subtraction are recorded and used to establish the geocentric velocity frame \citep[e.g.,][]{haf99, haf03}. Translation from the geocentric frame to LSR
is calculated using the same traditional value for the Standard
Solar Motion  adopted by 21 cm observers, +20 \kms\ toward
$\alpha_{1900}  = 18^{h} 0, \delta_{1900} = +30\fdg0$ \citep{haf03}.
A thorium argon (Th-Ar) lamp is used as the wavelength calibrator
for the \nii\ observations. Velocity calibration can also be
verified for the \nii\ observations by comparing \ha\ and \nii\
spectra from bright nebulae---with the assumption that the emission
from both lines takes on relatively similar velocity profiles
\citep{haf99}.

Our intensity calibration is tied to regular observations of NGC
7000 (North American Nebula; NAN), a large emission nebula in
Cygnus. NGC 7000 subtends a large enough solid angle to fill WHAM's
1\degree-beam and bright enough to achieve very high
signal-to-noise in our typical exposure times. We use the absolute
\ha\ intensity of NGC 7000 as determined by
\citet{sch81} who used the planetary nebula NGC 7662 and standard
stars as reference objects. This study measured the intensity of
the NGC 7000 to be $850\pm50$ R for a 49$^{\prime}$ FOV using a
Fabry-Perot spectrometer. \sii\ and \nii\ emission is calibrated by
applying a correction factor of 0.94 for \sii\ and 1.15 for \nii. These factors account for instrument
throughput differences between these lines and \ha\ throughout the
instrument \citep{haf99}.

\section{RESULTS AND DISCUSSION}
\label{sec:results}

\begin{figure*}[tbp]  
  \begin{center}
   \includegraphics[width=0.35\textwidth]{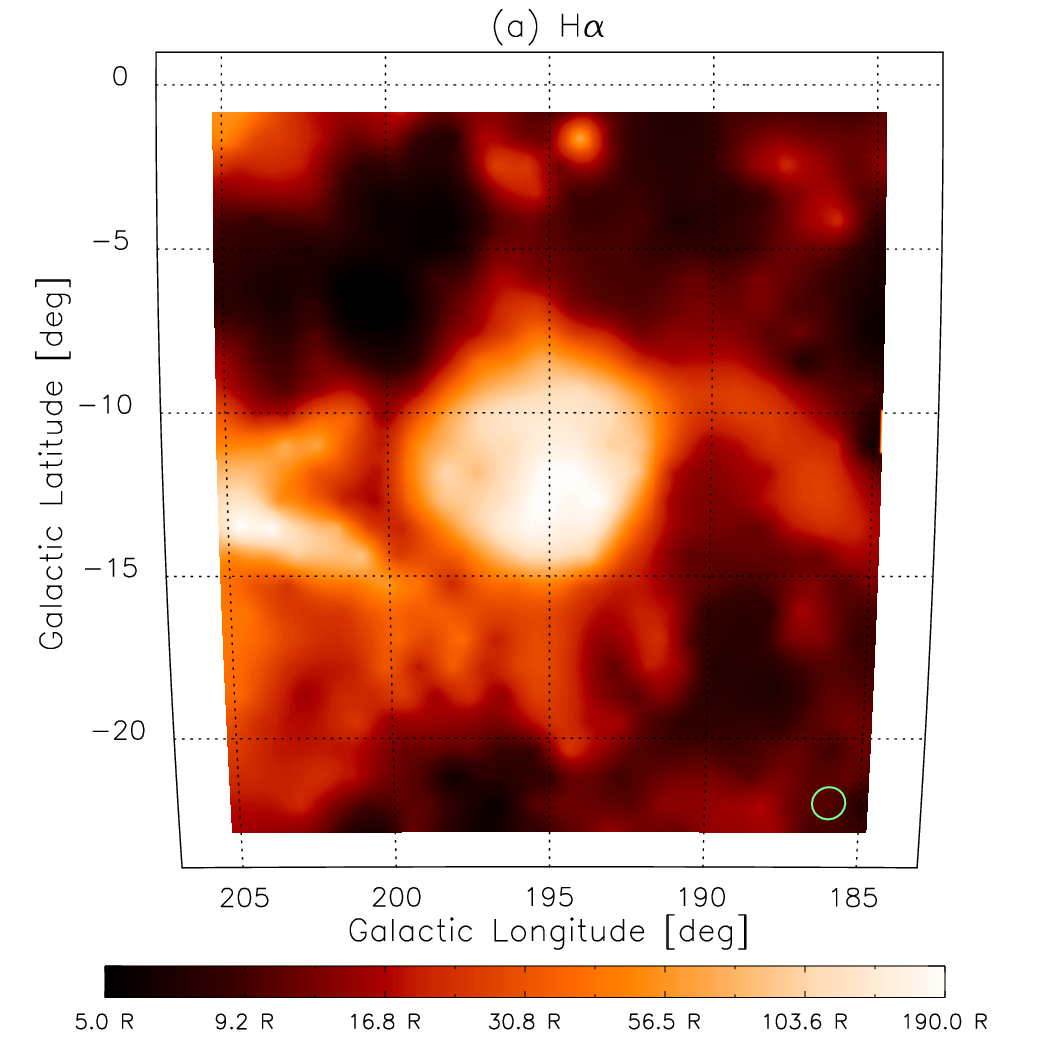} 
   \hspace{5ex}
   \includegraphics[width=0.35\textwidth]{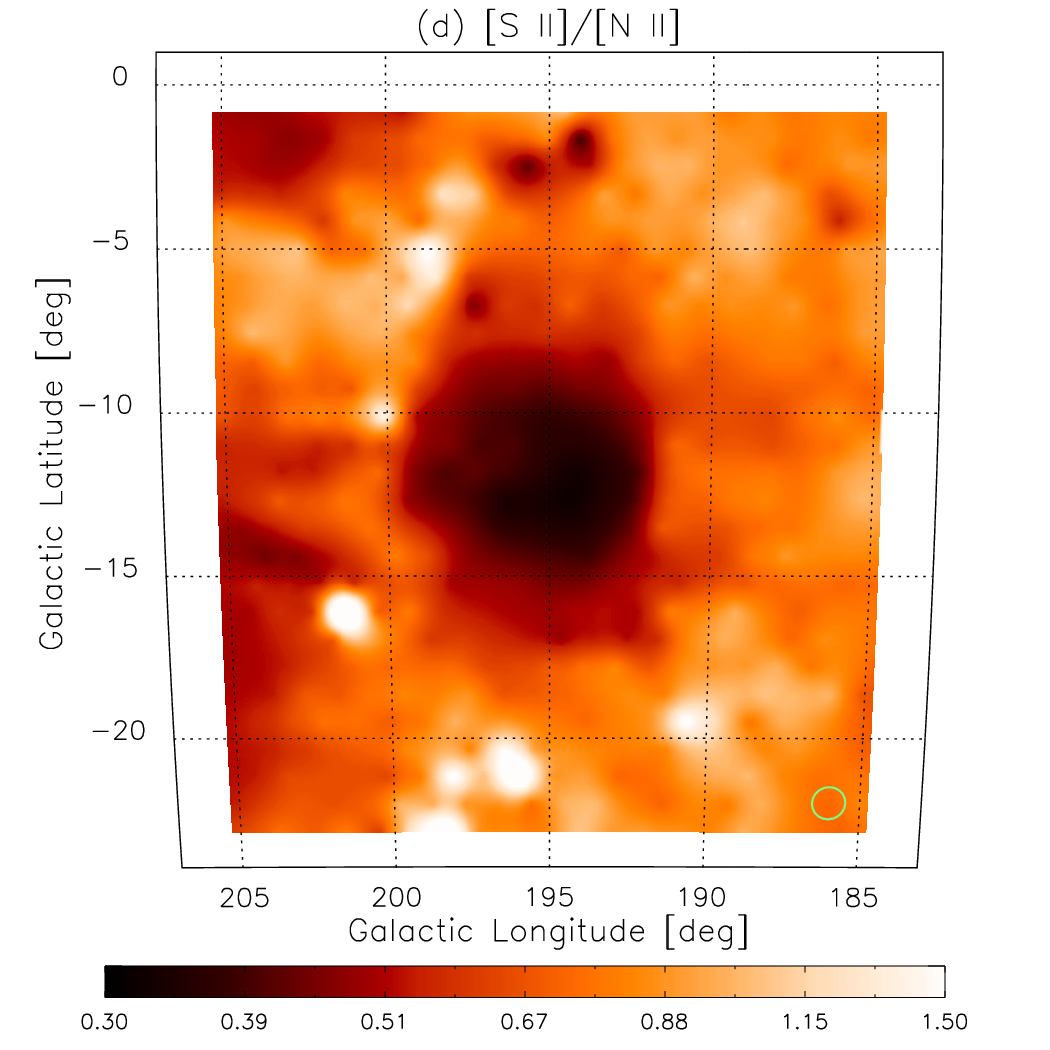}
   \includegraphics[width=0.35\textwidth]{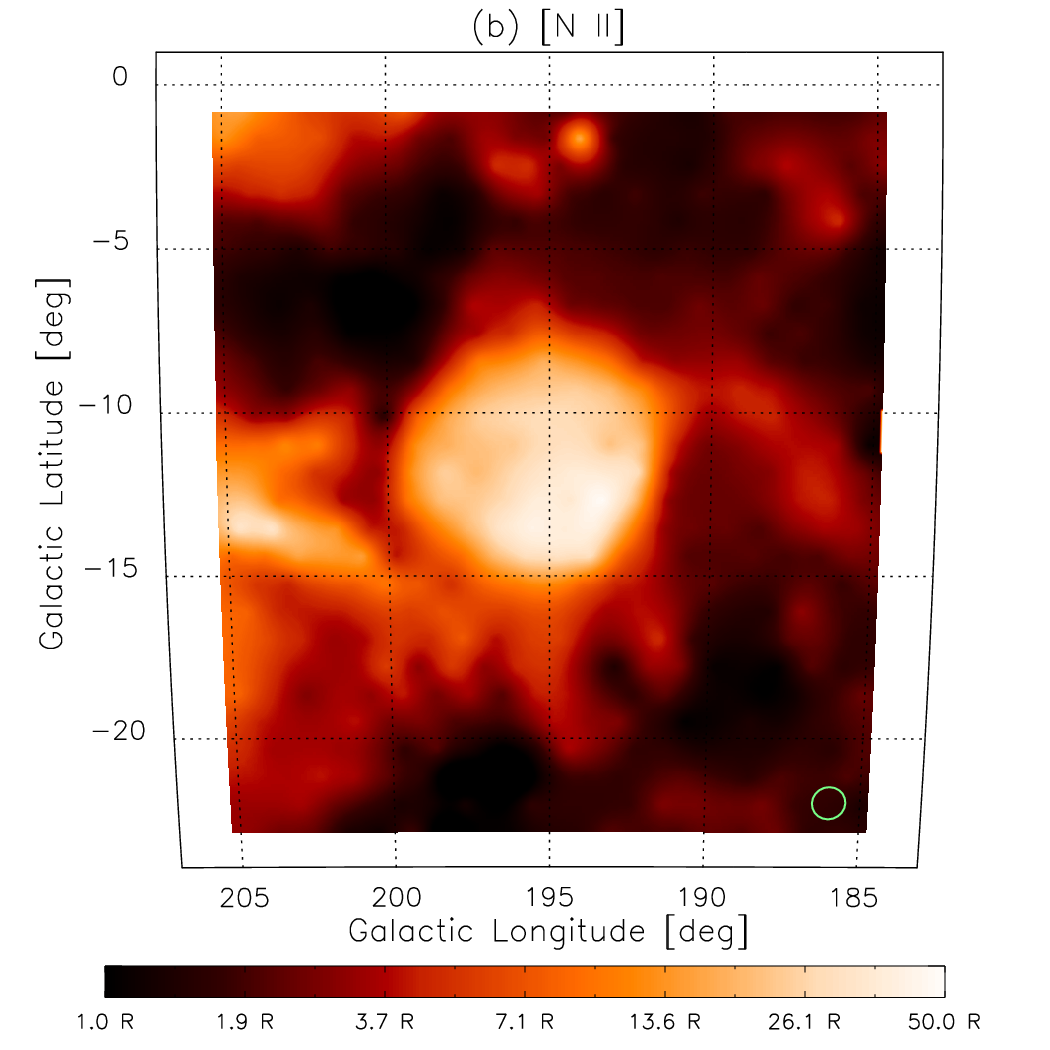}
   \hspace{5ex}
   \includegraphics[width=0.35\textwidth]{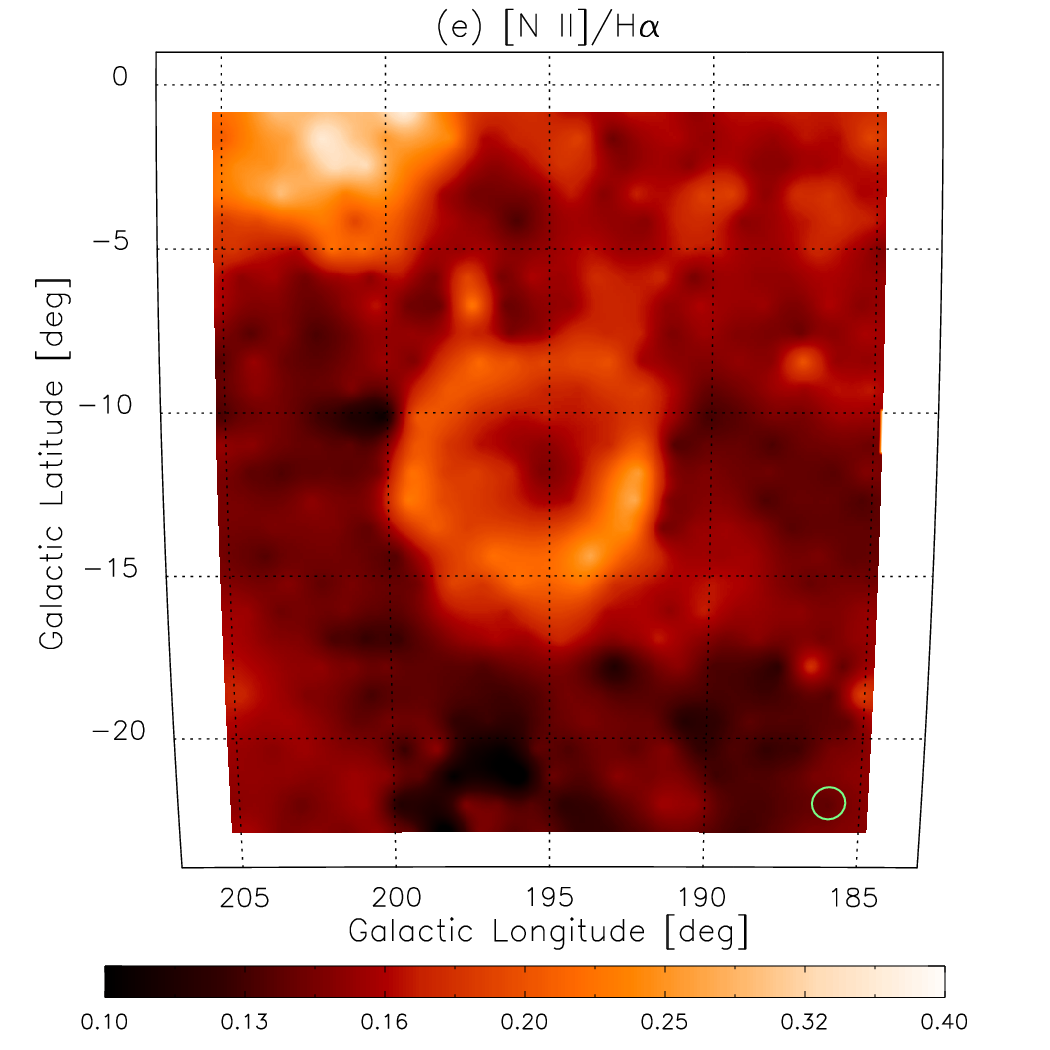}
   \includegraphics[width=0.35\textwidth]{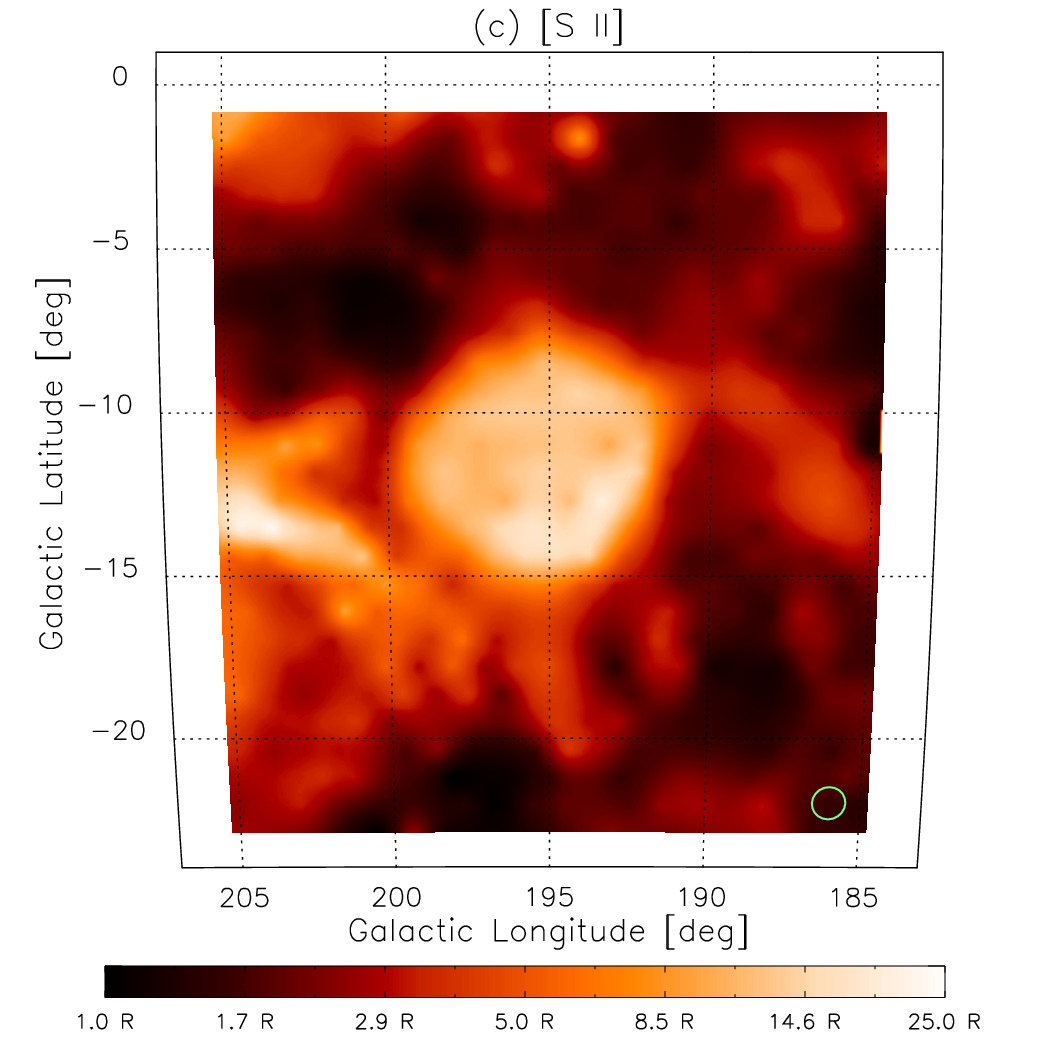}
   \hspace{5ex}
   \includegraphics[width=0.35\textwidth]{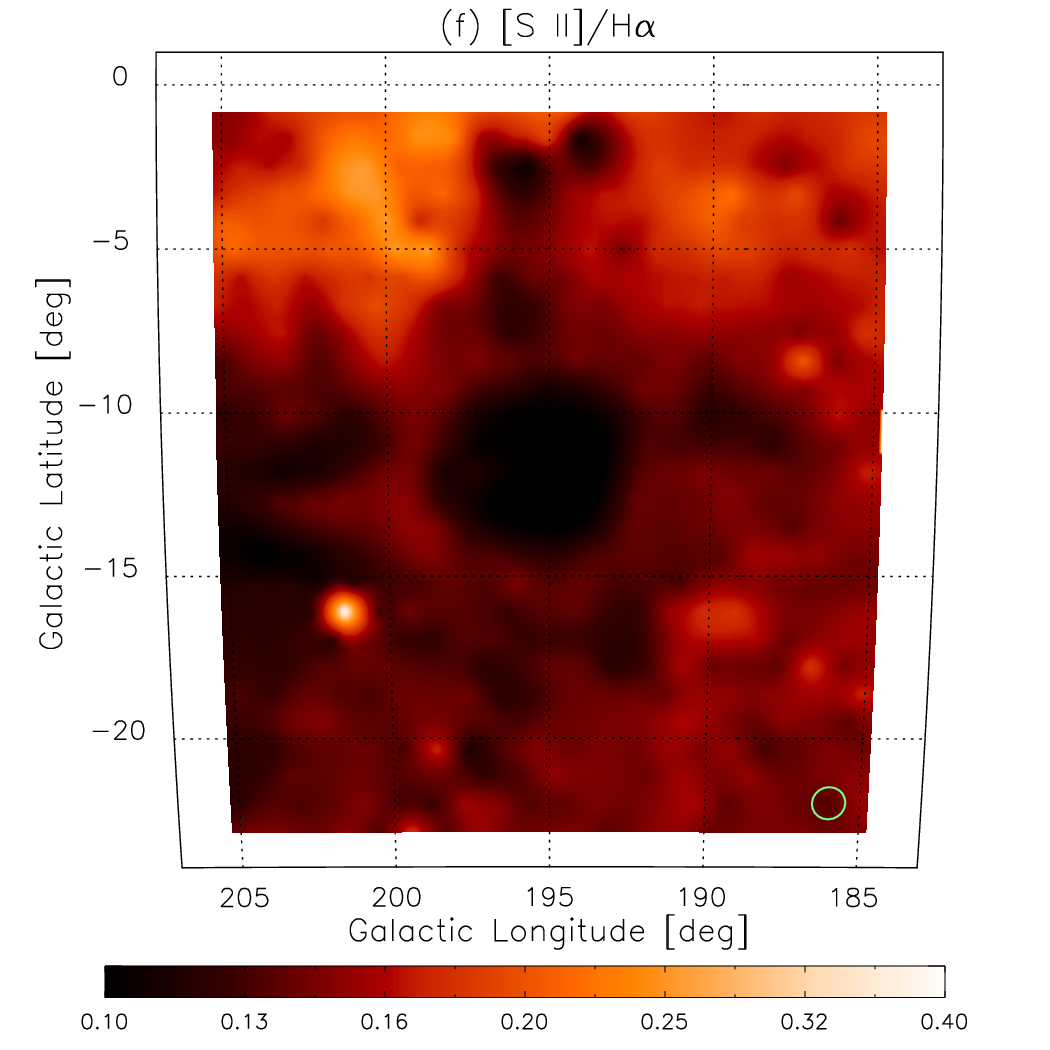}
    \caption{Smoothly interpolated \emph{a}) \ha, (\emph{b}) \nii, and (\emph{c}) \sii\ emission-line maps and (\emph{d}) \sii/\nii, (\emph{e}) \nii/\ha, and (\emph{f}) \sii/\ha\ line-ratio maps of the ionized region surrounding \lo\ $(\ell = 195\fdg05,\ b = -12\fdg00)$ as surveyed by WHAM. Intensity is integrated over velocities between $\vlsr = -80$ \kms\ and $+80$ \kms\ for each of the 547 individual spectra. Emission-line maps are logarithmically scaled and line-ratio maps are linearly scaled between the range denoted on the colorbar under each map. A circle in the bottom-right of each figure depicts the size of the WHAM beam on the sky. Axes are degrees of Galactic longitude and latitude.
 \label{fig:maps}}
  \end{center}
\end{figure*}

Figures~\ref{fig:maps}a, \ref{fig:maps}b, and \ref{fig:maps}c display intensity-integrated maps of \ha, \nii,
and \sii\ emission surrounding \lo\ as sampled in the WHAM survey
mode. Each map is constructed from 547 individual $1\degree$-beam
observations. The color represents the intensity of each emission
spectrum integrated over velocities $\vlsr = -80$ to $+80$ \kms\ 
after subtraction of atmospheric and continuum emission as discussed in \S\ref{sec:obs}. These three intensity maps look
qualitatively similar with the brightest emission around \lo\ at $(\ell, b) = (195\fdg05, -11\fdg99)$ revealing the $8\degree$-diameter Str\"{o}mgren sphere. Darker regions still contain emission from the
WIM but are much fainter than the \lo\ \hii\ region.  

Figures~\ref{fig:maps}d, \ref{fig:maps}e, and \ref{fig:maps}f show several line-ratio combinations: \sii/\nii, \nii/\ha, and \sii/\ha. These reveal that the \hii\ region has distinct spectral properties compared to the more diffuse surrounding emission---in addition to its elevated optical line intensities. The spatial distribution depicted in these maps complements the quantitative analysis in sections below that derive physical conditions of the gas surrounding \lo.


\begin{figure*}[tbp]  
  \begin{center}
  \plottwo{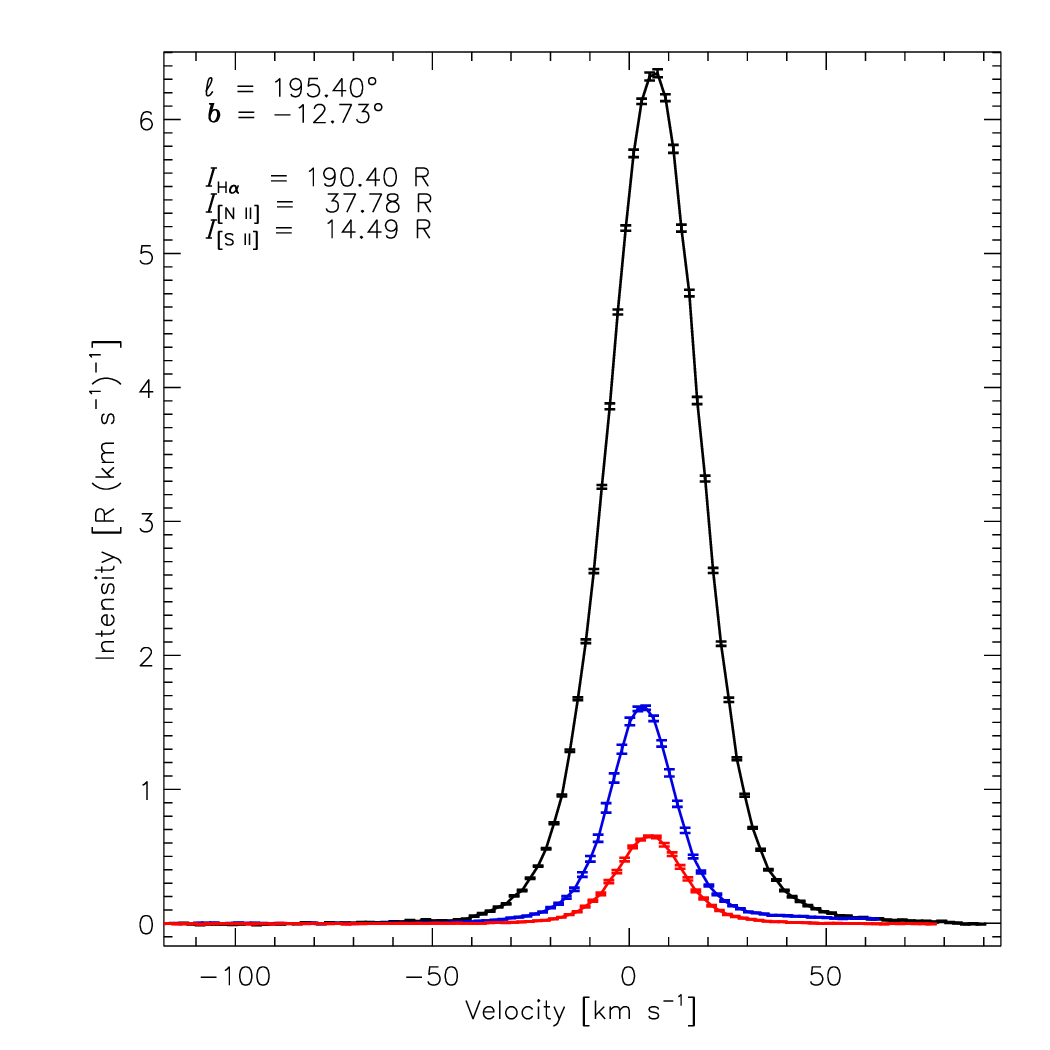}{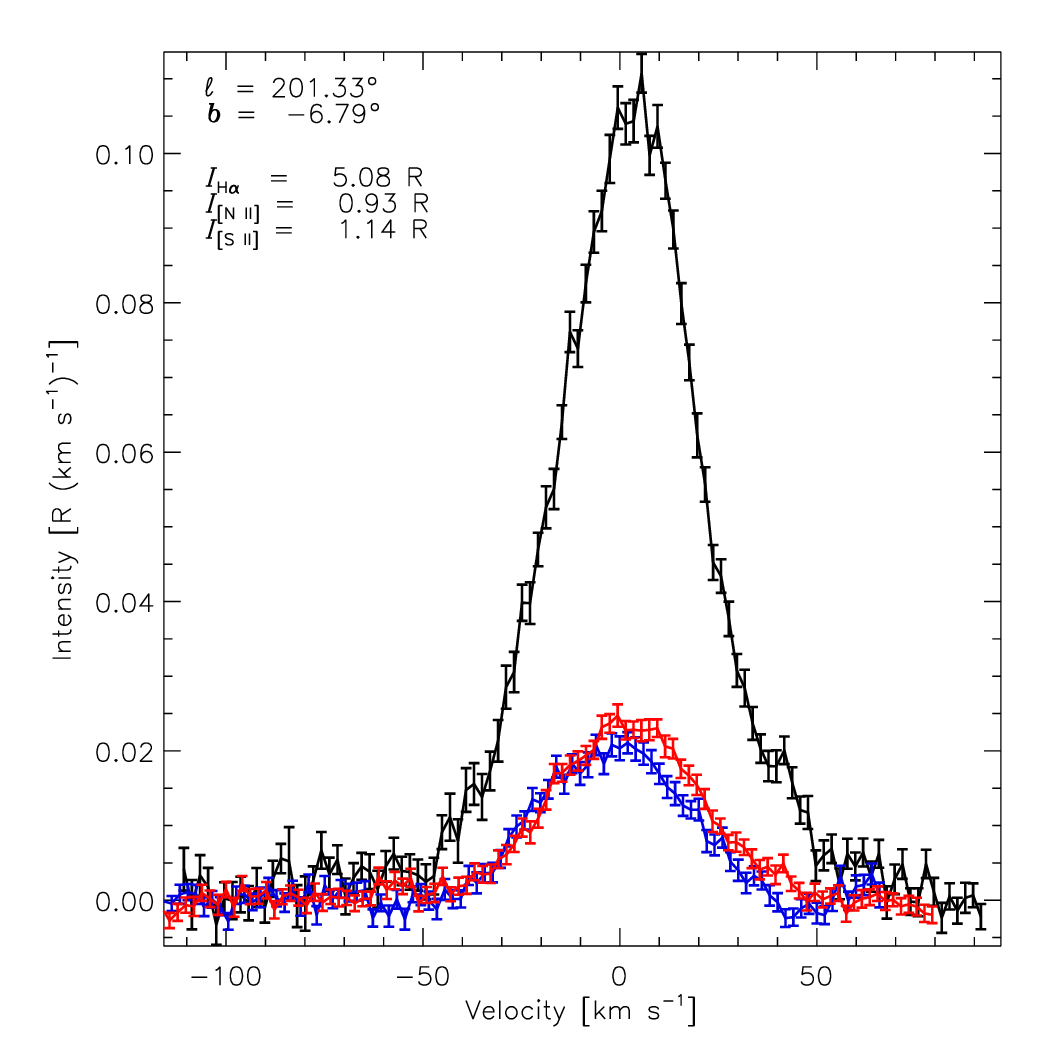}
  \caption{Sample \ha\ (\emph{black}), \nii\ (\emph{blue}), and \sii\ (\emph{red}) spectra from two directions, $(\ell,\ b) = (195\fdg4,\ -12\fdg73)$ and $(201\fdg33,\
-6\fdg79)$, are plotted as intensity vs.\ LSR velocity. Total integrated intensities for each line are noted in the upper left corner of each plot.  These two examples are selected from the 547 observations used to create the maps in Figure~\ref{fig:maps}.
 \label{fig:spectra}}
  \end{center}
\end{figure*}

Figure~\ref{fig:spectra} shows a pair of sample individual spectra in
each of the three lines along two different lines of sight in the region. The spectra highlighted in Figure~\ref{fig:spectra} are toward the brightest direction of the \hii\ region at $(\ell,\ b) = (195\fdg40,\ -12\fdg73)$
and one $\sim 8\degree$ away at $(\ell,\ b) = (201\fdg33,\
-6\fdg79)$. The spectra are plotted as intensity (R/(\kms)) versus LSR
velocity (\kms). The total integrated intensity of the emission lines
is labeled in the upper left corner of each plot. Note that the
intensity of \sii\ and \nii\ increases relative to \ha\ in fainter regions, as can also be seen in Figures~\ref{fig:maps}e and 1f. In sections below, we often concentrate on the
\hii\ region---observations within $4\degree$ of \lo. These bright spectra are dominated by the nebular emission and are fit well by a single Gaussian. This region of focus spans
roughly $\ell = 190\degree$ to $200\degree$ and $b = -17\degree$ to
$-7\degree$.


Table~\ref{tbl-1} summarizes several basic properties for a line of
sight centered near \lo. We
adopt the selective extinction parameter, $E(B - V)$, from
photometry and intrinsic colors presented in \citet{cru74};
spectral type from \citet{mai04}; and distance from
\citet{dol01}. Using this distance, we calculate the
distance away from the Galactic mid-plane, $z = d \sin b$, where $d$ is
the distance to \lo, and $b$ is the Galactic latitude. Beams with
elevated \ha\ extend to approximately $\theta_{H II} \sim 4\degree$,
which then leads to an estimate of the physical radius of the \hii\ region,
$R_{H II}$. The values for \lc\ and $n_e$ are computed
below in \S\ref{sec:ha}. Uncertainties on these last three parameters are dominated by the uncertainty in the distance to \lo.


\floattable
\begin{deluxetable*}{lcl}

\tabletypesize{\footnotesize}
\tablewidth{0pt}
\tablecaption{Selected Parameters of \lo\ A and Sharpless 2-264.\label{tbl-1}}

\tablehead{\colhead{Parameter} & \colhead{Symbol} & \colhead{Value}}

\startdata
Star & \ & HD 36861 \\
Spectral type & \  & O8 III \\
Galactic coordinates & $(\ell, b)$ & (195\fdg05, -12\fdg00) \\
Selective extinction & $E(B - V)$ & 0.12 \\
Distance from Sun & $d$ & $450 \pm 50$ pc \\
Distance from Galactic Plane & $z$ & $-94 \pm 10$ pc \\[1ex]
Angular radius of H {\sc ii} region & $\theta_{HII}$ & $4\arcdeg$ \\
Physical radius of H {\sc ii} region & $R_{HII}$ & $31 \pm 3$ pc \\
\ha\ intensity\tablenotemark{a} & \iha & $185 \pm 1$ R \\
Emission measure\tablenotemark{a} & EM & $579 \pm 3$ cm$^{-6}$ pc \\
Ionizing photon luminosity\tablenotemark{b} & $L_c$ & $(7.2^{+1.5}_{-1.7}) \times 10^{48}$ ph s$^{-1}$ \\
Average H {\sc ii} region density\tablenotemark{b} & $n_e$ & $ 2.5 \pm 0.3 $ cm$^{-3}$ \\
\enddata 

\tablenotetext{a}{As sampled by the one-degree WHAM beam toward \lo. Higher intensity emission may exist on smaller scales.}
\tablenotetext{b}{As derived in this work. See \S\ref{sec:ha}}

\end{deluxetable*}


\subsection{\ha\ EMISSION}
\label{sec:ha}

 From Haffner et al. (2003),
\begin{equation}
  EM \equiv \int n_{e}^{2}\,dl = 2.75\ T_{4}^{0.9}\ I_{H\alpha}\ (\textrm{cm}^{-6}\ \textrm{pc})
\label{eq:em-iha}
\end{equation}
where \iha\ is measured in Rayleighs. Including a correction for
dust extinction and assuming a temperature of 8000~K, we arrive at
this standard equation from \citet{rey82}:
\begin{equation}
  \textrm{EM} = 2.25\ \iha\ e^{2.75\ E(B-V)}\ (\textrm{cm}^{-6}\ \textrm{pc}).
\label{eq:em}
\end{equation}
For simplicity we assume the selective extinction, $E(B - V)$, is
the same throughout the \hii\ region as measured toward \lo. Future
work will incorporate \hb\ observations to correct \iha\ for each
WHAM beam. The peak \ha\ intensity (\iha) is measured to be 185 R.
From equation~(\ref{eq:em-iha}), we see that $\iha \propto \textrm{EM} =
\int n_{e}^{2}dl$, and thus the ionization of the nebula. By summing
over the total observed intensity of the \hii\ region, we can
estimate the luminosity of the Lyman continuum photons ($L_{c}$)
produced by \lo\ using the following equation (from
\citealt{haf01}):
\begin{equation}
  \lc ={\frac{4\pi d^{2}}{\epsilon}} \int \iha\ e^{\tau_{\ha}}\,d\Omega\ (\textrm{ph\ s}^{-1}),
  \label{eq:lc}
\end{equation}
where $d$ is the distance to the region, \iha\ is the \ha\ intensity
emitted over the solid angle $d\Omega$, and $e^{\tau_{\ha}}$ is
the extinction correction, defined observationally with $e^{2.75\
E(B-V)}$. The factor 2.75 is computed from the extinction models of
\citet{car89} at \ha\ assuming the typical diffuse total to
selective extinction ($R_v$) ratio of 3.1. ${\epsilon}$ is the
fraction of \ha\ photons produced per Lyman continuum photon, 0.47
for gas at 8000~K \citep{mar88}. We compute the \ha\ intensity of the \hii\ region using $\iha\ = (I_{\mathrm{H\alpha,obs}} - I_{\mathrm{H\alpha,bf}})$. Here, $I_{\mathrm{H\alpha,obs}}$ is the observed, integrated intensity ($\vlsr = -80$ to $+80$ \kms) of each spectrum, and $I_{\mathrm{H\alpha,bf}}$ is the intensity along the line of sight from behind and in front of the \hii\ region. Examining the distribution of intensities in this region by taking cuts through Sh~2-264, we estimate the average background/foreground emission $I_{\mathrm{H\alpha,bf}} = 10$ R. 

Approximating the integral in equation~(\ref{eq:lc}) as a sum over WHAM beams,
\begin{equation}
  \lc = 2.03\times10^{43}d^{2}\left({\frac{\Omega_{HII} }{N\Omega_{p}}}\right) \\
  \sum_{N}(I_{\mathrm{H\alpha,obs}} - I_{\mathrm{H\alpha,bf}})\ e^{2.75\ E(B-V)}\ \Omega_{p}\   ( \textrm{ph\ s}^{-1}).
\end{equation}
Here, $\Omega_{p}$ represents the solid angle subtended by a  single
pointing in the 1\degree-circular beam of WHAM ($2.39 \times
10^{-4}$ sr); $\Omega_{H II}$/(N$\Omega_{p}$) is a
correction factor that estimates gaps and overlaps of the survey
grid within the \hii\ region; and $N$ is the number of observations within
a $4\degree$-radius region surrounding \lo. For a circular
region on the sky (i.e., cone in solid angle),
\begin{equation}
  \Omega_{H II} = 2 \pi (1 - \cos (4\degree)) = 1.53\times 10^{-2}\ (\textrm{sr})
\end{equation}
and
\begin{equation}
  \Omega_{HII}/N\Omega_{p} = (1.53 \times 10^{-2}\ \textrm{sr})/(60 \times 2.39\times10^{-4}\ \textrm{sr}) = 1.07.
\end{equation}

Using the $E(B - V)$ and distance in  Table~\ref{tbl-1}, we find $\lc =
(7.2^{+1.5}_{-1.7}) \times 10^{48}$ ph s$^{-1}$. The uncertainty on this value is dominated by the
uncertainty of the distance to \lo\ resulting in a $\sim20\%$ on our estimate of
\lc. Our calculation of \lc\ is low compared to recent model
calculations for O8 III stars, $\sim$1.7 $\times$ 10$^{49}$ ph
s$^{-1}$ \citep[e.g.,][]{vac96}. However---assuming the model
reflects the true output of the star---our estimate using the \ha\
intensity of the \hii\ region is expected to be a lower limit if any
ionizing flux escapes the region. Our correction for \ha\ removed from our line of sight by dust scattering ($e^{\tau_{\ha}}$) using the selective extinction toward \lo\ assumes the value is constant across the nebula. We have no specific reason to expect this correction to be too low on average, but future work using \hb\ will remove this assumption. 

With \lc, we can also estimate the density ($n_{e}$) of the emitting
gas. In a spherical \hii\ region where H in the nebula is assumed to
be fully ionized and $n_{e}$ constant, the recombination balance
equation is:
\begin{equation}
\lc = {\alpha_{B}}n_{e}^{2}{\frac{4\pi}{3}}R_{HII}^{3}
\end{equation}
where, $\alpha_{B}$ is the recombination coefficient and equals to
3.10 $\times$ 10$^{13}$ cm$^{3}$ s$^{-1}$ for gas at $T \sim 8000$~K
\citep{ost89}. Then,
\begin{equation}
n_{e} = 162\ \sqrt{\frac{L_{c48}}{R_{HII}^{3}     }     }\ \
(\textrm{cm}^{-3})
\end{equation}
In this equation,  $L_{c48}$ is in units of $10^{48}$ ph s$^{-1}$,
$R_{H II}$ is in pc, and $n_{e}$ is in cm$^{-3}$. With R$ _{H II} = 31\pm3$ pc and $L_{c48} = 7.2^{+1.5}_{-1.7}$ we find $n_e = 2.5 \pm 0.3$ cm$^{-3}$.

\subsection{\sii\ AND \nii\ LINE RATIOS}\
\label{sec:ratios}

The relative intensities of metal-line emission to \ha\ and each
other can reveal information about the ionization state and
temperature of diffuse ionized gas
\citep[e.g.,][]{haf99,rey99,mad06}. We use
the \ha, \sii, and \nii\ intensity maps shown in
Figure~\ref{fig:maps} to examine
\nii/\ha, \sii/\ha, and \sii/\nii\ emission line ratio behavior. The \ha\ intensity
decreases by a factor of approximately 19 from the center
to the edge of the \hii\ region (with our one-degree beam sampling),
from $\iha \sim 190$ R near the center to $\iha \sim 10$ R on the
periphery of the area under analysis. However, line ratios vary by much less. The \sii/\ha\ ratio increases
from $\sim 0.1$ at the center to $\sim 0.2$ at the edge, and the
\nii/\ha\ ratio increases from $\sim 0.2$ at the center to $\sim
0.35$ at the edge of the nebula, factors of 2 and 1.75 from the center
to the edge of the nebula, respectively. The \sii/\nii\  ratio
also increases by about a factor of 2.5 from $\sim 0.4$ at the
center to $\sim 1$ at the edge. The range of in the \sii/\ha, \nii/\ha, and \sii/\nii\ line ratios
for this diffuse nebula are not too different from those seen in the
WIM \citep[e.g.,][]{mad06}, but the trends show different behaviors. 

Past studies
of diffuse \ha, \sii, and \nii\ and more recent investigations of the WIM
carried out with WHAM show that there is a good correlation between
the line ratios and variation in the temperature of the WIM
\citep{rey85a,rey88,haf99,rey99}. \citet{haf99} suggested that the tight correlation of \sii/\nii\
seen in their data arose simply from the dominance of S$^{+}$ and
N$^{+}$ ionization states in this gas together with similar
excitation energy of these two emission lines. They derived
estimates of the electron temperature and ionization state of sulfur
for a large region of the sky. While all their assumptions for the
WIM are not valid in a classical \hii\ region since the ionization
structure of emission nebulae changes significantly over smaller
scales, the techniques used to examine the line ratio behavior are
still useful here.

The basic equation for the intensity of emission lines from
collisionally excited ions found in \citet{ost89} and \citet{haf99}
is given as:
\begin{equation}
  I_{\nu} ={\frac{f_{\nu}}{4\pi}} \int
   n_{i}n_{e}{\frac{8.63\times10^{-6}}{T^{0.5}}}{\frac{\Omega(i,j)}{w_{i}}}e^{(E_{i,j}/kT)}dl
\label{eq:excite}
\end{equation}
where $I_{\nu}$ is in units of ph s$^{-1}$ cm$^{-2}$ sr$^{-1}$,
$\Omega(i,j)$ is  the collision strength of the transition,
$\textit{w$ _{i}$}$ is the statistical weight of the ground level,
$\textit{E$ _{ij}$}$ is the energy of the upper level of the
transition above the lower, $\textit{n$ _{i}$}$ and $\textit{n$
_{e}$}$ is the ion and electron density, respectively, $\textit{k}$
is Boltzmann's constant, $\textit{T}$ is the temperature of the gas,
$\textit{dl}$ is the pathlength through the emitting region, and
$f_{\nu}$ is the fraction of downward transitions that produce the
emission line in question  \citep[e.g.,][]{ost89,haf99}.

The intensities of \sii\ $\lambda$6716 and \nii\ $\lambda$6583
emission are then:
\begin{equation}
  I_{6716}(R)=2.79\times10^{5}  \left({\frac{H^{+}}{H}}\right)^{-1}\left({\frac{S}{H}}\right)\left({\frac{S^{+}}{S}}\right)T_{4}^{-0.593}
  \,e^{-2.14/T_{4}}\ \textrm{EM},
\label{eq:sii}
\end{equation}
and
\begin{equation}
  I_{6583}(R)=5.95\times10^{4}  \left({\frac{H^{+}}{H}}\right)^{-1}\left({\frac{N}{H}}\right)\left({\frac{N^{+}}{N}}\right)T_{4}^{-0.474}
  \,e^{-2.18/T_{4}}\ \textrm{EM},
\label{eq:nii}
\end{equation}
where  $T_{4}$  is in units of 10$^{4}$ K and EM is related to \iha\
in equation~(\ref{eq:em-iha}). The ratio of the metal emission lines is
then:
\begin{equation}
 {\frac{ I_{6716}}{I_{6583}}}  =4.69\,{\frac{({\frac{S}{H})}}{{(\frac{N}{H})}}}  {\frac{({\frac{S^{+}}{S})}}{{(\frac{N^{+}}{N})}}}\,
 T_{4}^{-0.119}\,e^{0.04/T_{4}}.
 \label{eq:siinii}
\end{equation}
 \citet{haf99} adopted parameters for (S/H)$= 1.86 \times
10^{-5}$ from  Anders and Grevesse (1989) and (N/H)$ = 7.5 \times
10^{-5}$ from Meyer et al. (1997) for the gas-phase abundances of S
and N. This ratio is insensitive to temperature because of the
nearly identical energies required to excite the lines. As a result,
when \sii/\nii\ is seen to be constant throughout a spatial region,
$({S^{+}}/{S})/({N^{+}}/{N})$ also changes little.

 Models  \citep{sok94,sem00}  have shown in diffusely ionized regions (e.g.,
the WIM, IVCs, HVCs, etc.) N$^{+}$/N$^{0}$ tracks H$^{+}$/H$^{0}$
because of their similar first ionization potentials (14.5 eV and
13.6 eV, respectively) and a weak charge-exchange reaction. This
relationship allows us to use equation~(\ref{eq:nii}) as a rough estimate
of the electron temperature within the emitting gas. Furthermore, in
these regions where the fractional ionization of H has been also
measured to be near unity \citep{rey98,mad06}  and higher ionization
species such as He$^{+}$ and O$^{++}$ are extremely faint or not
detected \citep{rey95,tuf97,how99,mad06}, N$^{+}$ should be the
dominant ion state with N$^{+} \sim 0.8$ -- 1.0 \citep{haf99}. On
the other hand, S has a second ionization potential of 23.4 eV, just
below neutral He at 24.6 eV. As a result, it is likely partially
ionized to S$^{++}$. Using this justification and
equation~(\ref{eq:siinii}), \citet{haf99}  estimated ionization states of
S$^{+}$/S that range between about 0.25 and 0.8 in the WIM, with an
average in the local diffuse background of 0.6 -- 0.65. Such ratios
have been found to be systematically higher in the WIM than in
classical \hii\ regions \citep[e.g.,][]{rey88,rey99,haf99}. The
lower S$^{+}$/S ratios in \hii\ regions is somewhat expected since
the ionization parameter ($n_{ph}/n_e$) is large and S$^{+}$ is
ionized to S$^{++}$ more readily, which produces a corresponding
decrease in the observed \sii\ intensity relative to \ha\
\citep{dom94,sok94,haf99}. However, \citet{haf99}  show that a
higher electron temperature ($T_{e}$) in the WIM will also increase
the \sii/\ha\ ratio.


\begin{figure*}[tbp]  
  \begin{center}
	\plottwo{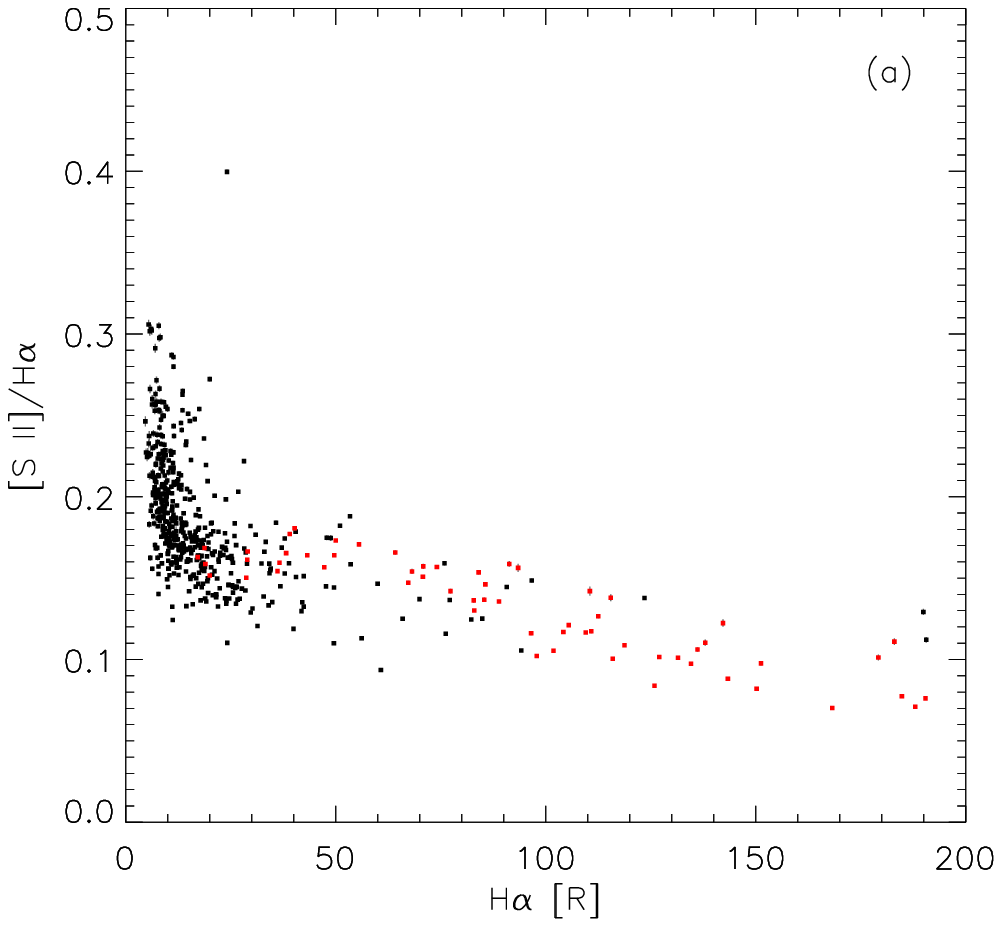}{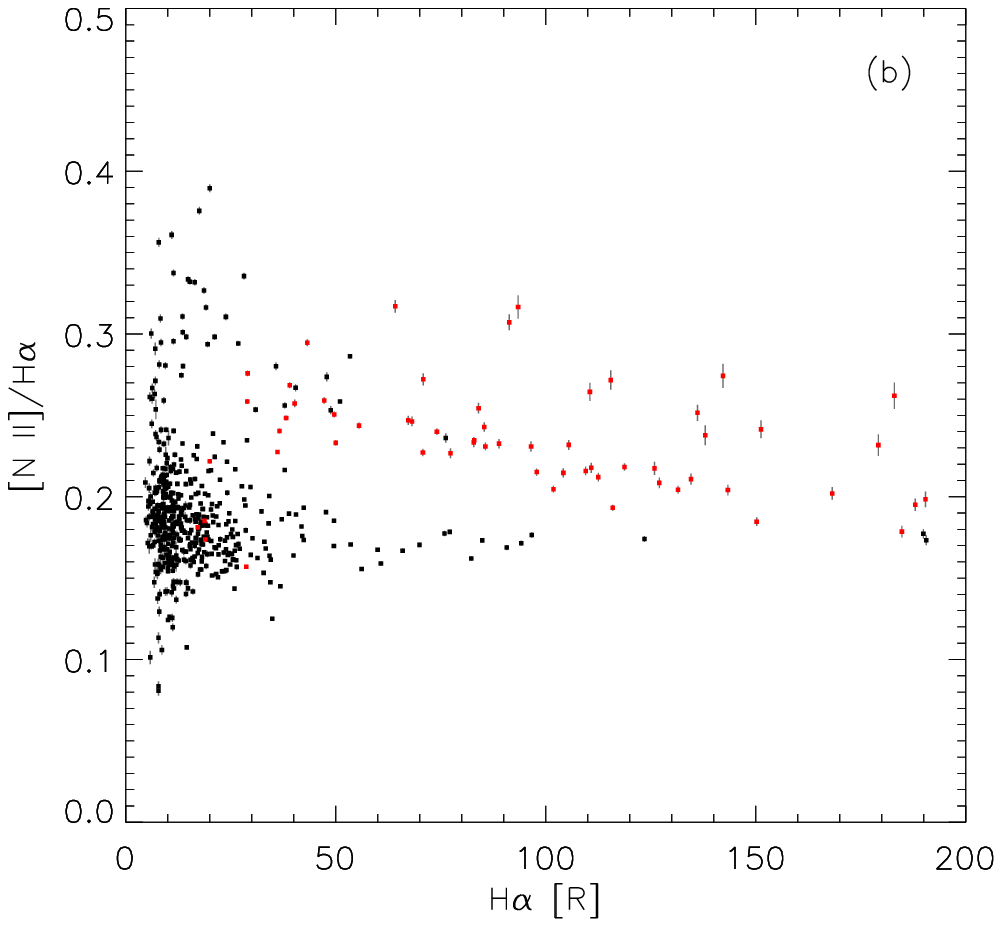}
	\plottwo{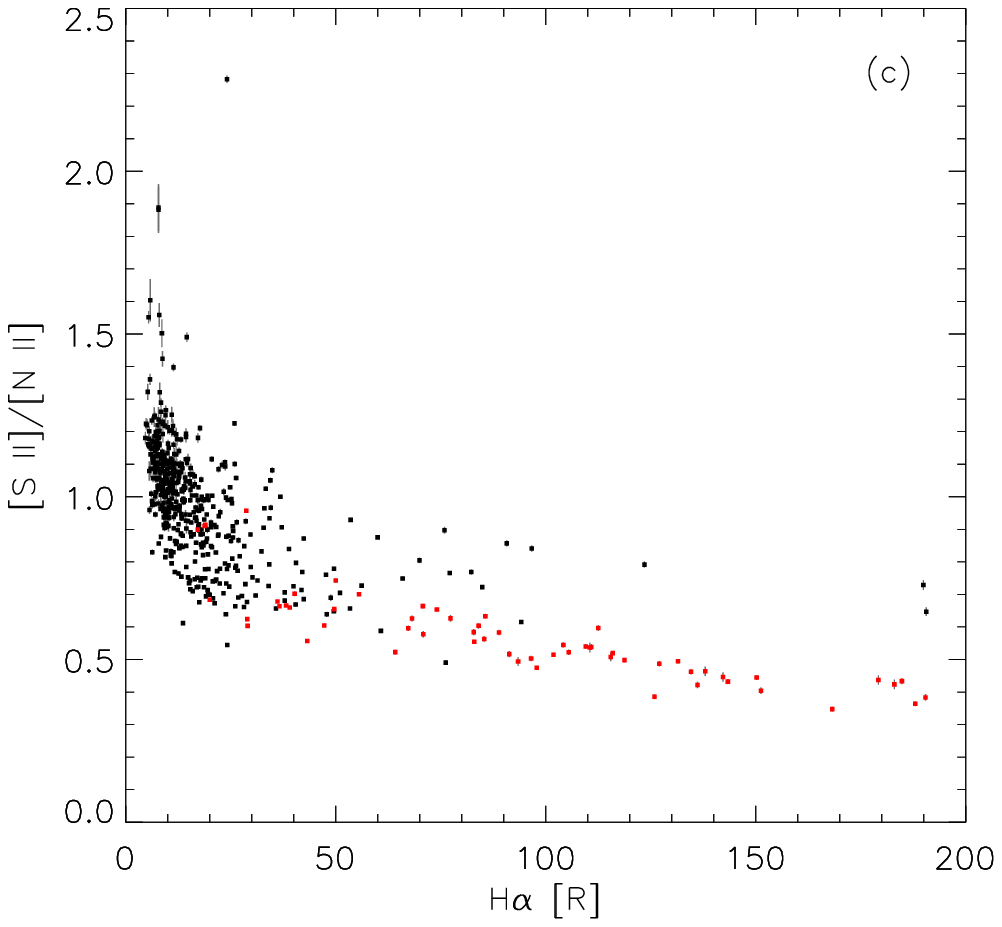}{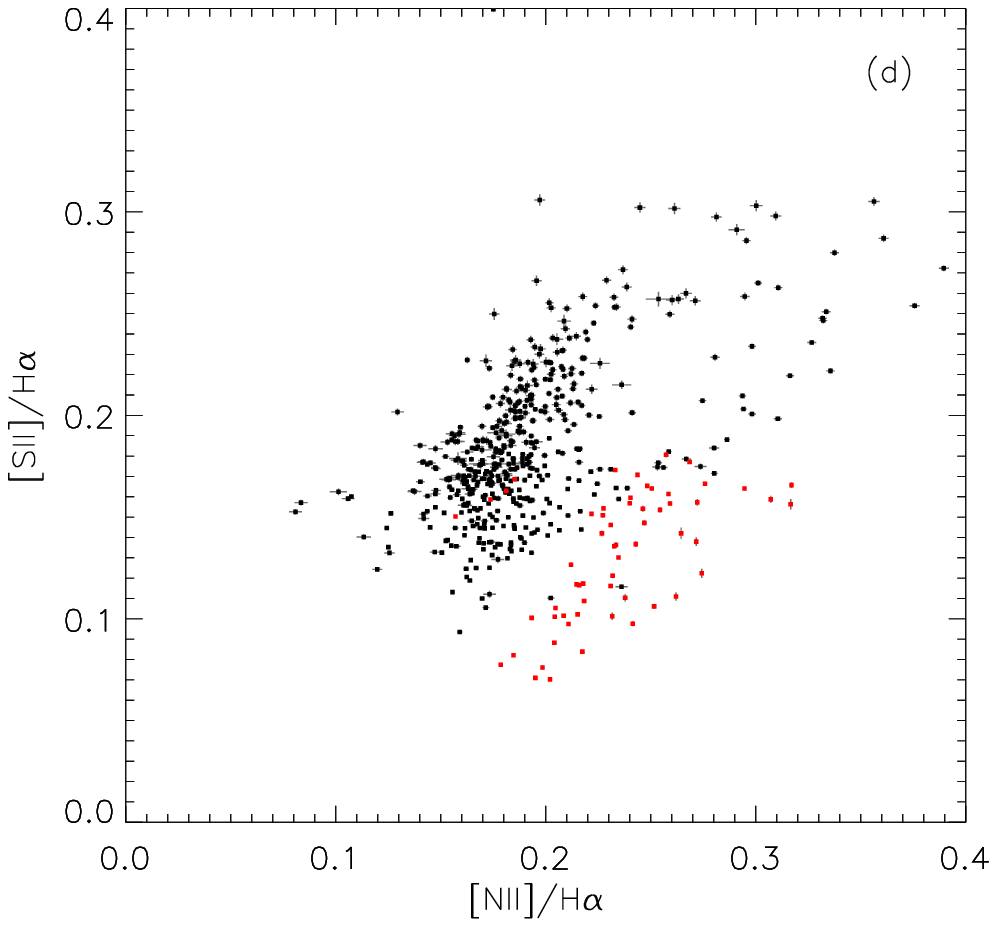}
    \caption{ (\emph{a}) \sii/\ha\ vs.\ \ha,  (\emph{b}) \nii/\ha\ vs.\ \ha,  (\emph{c}) \sii/\nii\ vs.\ \ha, and  (\emph{d}) \sii/\ha\ vs. \nii/\ha\ relationships for pointings in Figure~\ref{fig:maps} around \lo. Intensities are integrated between $\vlsr = -80$ and $+80$ \kms. Error bars (\emph{gray}) on each symbol represent the uncertainty of the intensity or ratio. Many error bars are smaller than the plotted symbols since uncertainties on integrated intensities are very small for WHAM observations in this moderately bright region of the Galaxy. Data points within 4\arcdeg\ of \lo\ are highlighted in red. 
    \label{fig:ratios}}
  \end{center}
\end{figure*}


Figure~\ref{fig:ratios} shows the behavior of \sii/\ha, \nii/\ha\, and \sii/\nii\ as a function of \ha\ intensity as well as \sii/\ha\ vs.\ \nii/\ha\ throughout this region containing Sh~2-264. The core of the \hii\ region, observations within 4\arcdeg\ of \lo, are highlighted in red. In general, Figure~\ref{fig:ratios}a shows a marked increase in \sii/\ha\ ratio with decreasing \ha\ intensity, both inside and outside the \hii\ region. A trend in the \nii/\ha\ ratio in Figure~\ref{fig:ratios}b is less pronounced, although the \hii\ region has a distinctly elevated track in this ratio, which is visible as the radial gradient close to the star in Figure~\ref{fig:maps}e.
As equation~(\ref{eq:siinii}) shows, Figure~\ref{fig:ratios}c  (\sii/\nii\ vs.\ \ha) is fairly insensitive to temperature so that variation is dominated by the ion ratio (S$^+$/N$^+$). Figure~\ref{fig:ratios}d
plots \sii/\ha\ and \nii/\ha\ against one other, with the \hii\ region again occupying a distinct area of the figure. The distinct separation of \hii\ regions from the more diffuse background is common in these studies  (\emph{e.g.}, Figures 7 \& 13 of \citealt{mad06}) and highlights the different natures of the radiation field near ionizing sources and out in the diffusely ionized and more widely distributed ISM.

\subsection{LINE WIDTHS}
\label{sec:widths}

We can use the widths (FWHM) of \ha\ and \sii\ emission lines to
measure temperature and nonthermal velocity throughout the \lo\
\hii\ region. This method was utilized by \citet{rey85a}   to
show that the properties of diffuse gas along several lines of
sights in the WIM differs from those in classical \hii\ regions. The temperature
and nonthermal velocity of the emitting gas in Sh~2-264 can be derived
from the total line widths of \sii\ and \ha, if the emission lines
originate from ions that are uniformly mixed within the \hii\ region
\citep{rey85b,rey88}.

\begin{figure}[tbp]  
  \begin{center}
 \plotone{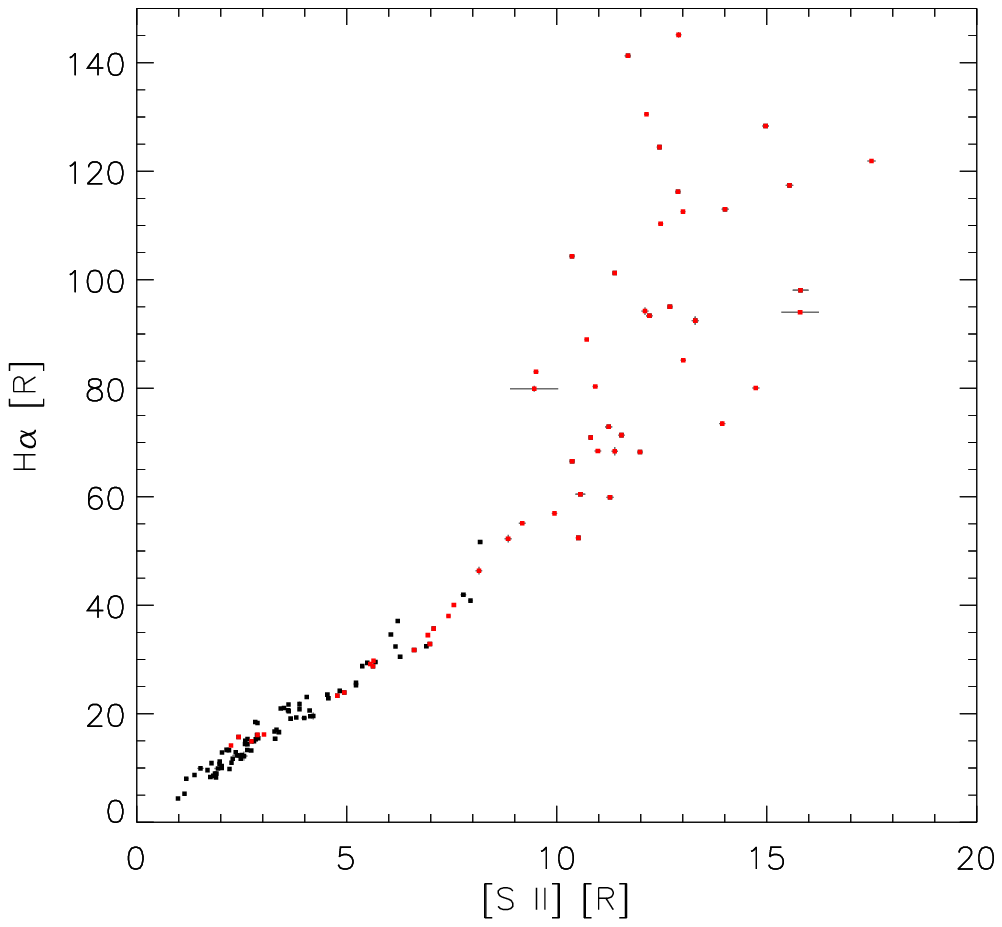}
 \caption{\ha\  vs. \sii\ fitted component areas for observations within 6\arcdeg\ of \lo, the rough extent of its \hii\ region. Those pointings closer than 4\arcdeg\ are highlighted in red.  Error bars (\emph{gray}) on each symbol represent a 1$\sigma$ uncertainty in the fitted Gaussian component area. Note that many error bars are smaller than the plotted symbols since the area (intensity) is very well-determined by WHAM observations in this moderately bright region of the Galaxy. Uncertainties do increase in some directions when spectral profiles become more complex and require several comparable Gaussian components. 
  \label{fig:icorr}}
  \end{center}
\end{figure}

\begin{figure}[tbp]
\begin{center}
  \plotone{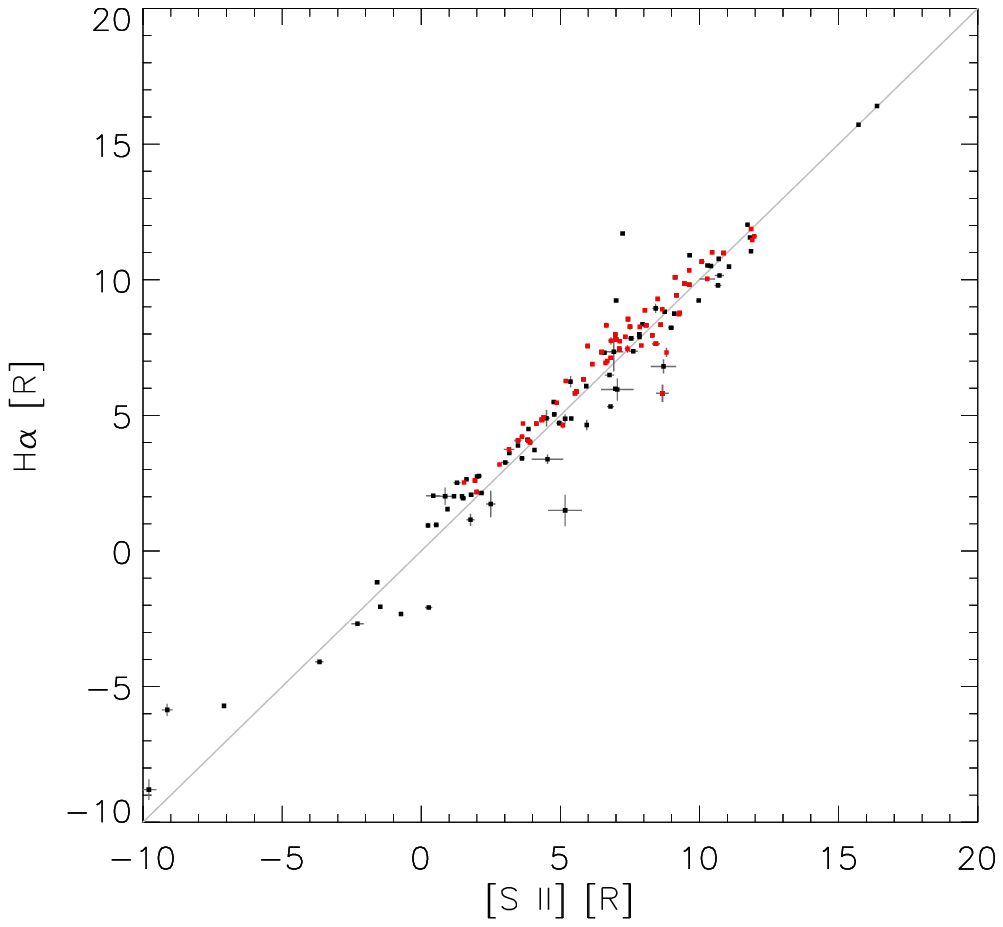}
\caption{\ha\  vs. \sii\ fitted component means for observations within 6\arcdeg\ of \lo, the rough extent of its \hii\ region. Those pointings closer than 4\arcdeg\ are highlighted in red. Error bars (\emph{gray}) on each symbol represent a $1\sigma$ uncertainty in the fitted Gaussian mean. A line of unity (\emph{gray}) is provided for reference.
\label{fig:vcorr}}
\end{center}
\end{figure}

Some evidence that the ions share a kinematic profile is provided by
Figure~\ref{fig:spectra} above, which shows the agreement in the
shape and extent of the line profiles of \sii\ and \ha\ emission.
Further confidence comes from comparing some of the quantitative
properties of the line emissions. In Figure~\ref{fig:icorr}, we plot
the intensity of the \ha\ emission against the intensity of
corresponding \sii\ emission. This figure reveals that the \sii\ is
generally correlated with \ha\ (deviations are highlighted in the
\sii/\ha\ ratio plot of Figure~\ref{fig:ratios}a). Furthermore, the
peak of these lines track each other in radial velocity as seen in
Figure~\ref{fig:vcorr}. The correlation in the plots of Figures
\ref{fig:icorr} and \ref{fig:vcorr} as well as the similarity in
shape of the line profiles in Figure~\ref{fig:spectra} provide
sufficient evidence that components within the two lines originate
from the same region of gas for the purpose of our analysis.
However, our ability to accurately derive the widths decreases when
the \hii\ region component becomes comparable to the foreground
and/or background emission in this complex area of the Galaxy. For
the rest of this analysis, we use only the observations
within a 4\degree-radius of \lo\ where we are able to isolate the \hii\ region emission.


The \ha\ and \sii\ total line widths (FWHM), $W_{H}$ and $W_{S}$,
are given by:
\begin{equation}
W_{H}\approx21.4\left[{\frac{T_i}{10^{4}}}+\left({\frac{v}{12.8}}\right)^{2}+0.070\right]^{1/2}
(\textrm{km s}^{-1})
 \label{eq:wh}
\end{equation}
and
\begin{equation}
W_{S}\approx3.78\left[{\frac{T_i}{10^{4}}}+\left({\frac{v}{2.27}}\right)^{2}\right]^{1/2}
(\textrm{km s}^{-1})
 \label{eq:ws}
\end{equation}
where $T_i$ is temperature as measured by the ion speeds and $v$ is
the most probable speed of the nonthermal velocity distribution; the
extra 0.070 term in equation~(\ref{eq:wh}) accounts for width due to the
fine structure of the \ha\ line  \citep{rey85a}. It is reasonable to
treat the nonthermal radial velocity distribution as a Gaussian
because the \sii\ components, whose widths are dominated by
nonthermal motions at these temperatures (see equation~(\ref{eq:ws}) and
the limit on $T_i$ and $v$) are well-fit by Gaussian profiles.

The mode of the nonthermal velocity distribution, or most probable
speed $v$, and the gas temperature $T_i$ are then given by:
\begin{equation}
v\approx \left[ {\frac{\left({\frac{W_{H}}{21.4}}\right)^{2}-
\left({\frac{W_{S}}{3.78}}\right)^{2}-0.70}
     {\left({\frac{1}{12.8^{2} }}-{\frac{1}{2.27^{2} }}\right)
}}\right] ^{1/2} (\textrm{\kms})
 \label{eq:vnt}
\end{equation}
\begin{equation}
T_i\approx 10^{4}\left[ \left({\frac{W_{S}}{3.78}}\right)^{2}-
\left({\frac{v}{2.27}}\right)^{2}   \right]
 =10^{4}\left[ \left({\frac{W_{H}}{21.4}}\right)^{2}-
\left({\frac{v}{12.8}}\right)^{2}   -0.070\right] (\textrm{K})
 \label{eq:tion}
\end{equation}


\begin{figure*}[tbp]  
  \begin{center}
  \plotone{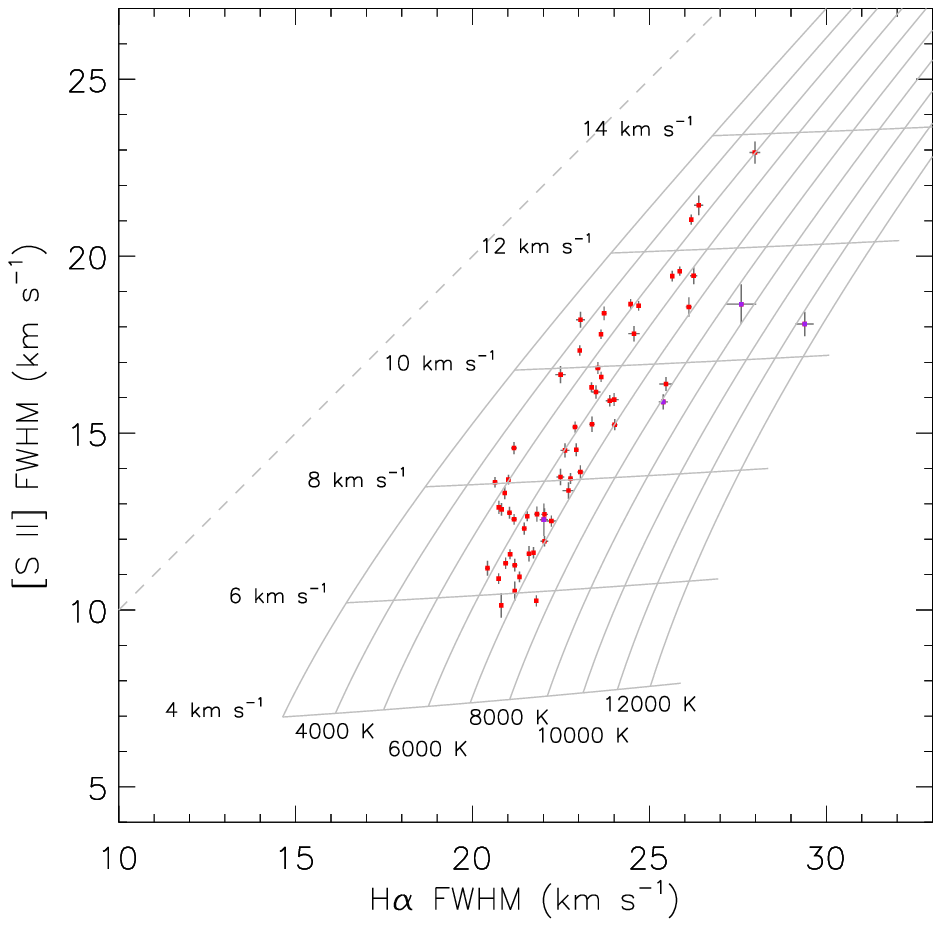}
 \caption{Width (FWHM) of \sii\ emission components plotted against the width of their corresponding \ha\ component. Error bars (\emph{gray}) represent a $1\sigma$ uncertainty of the fitted Gaussian widths. The solid, gray grid lines indicate the combination of temperature and nonthermal gas motion that accounts for each pair of \sii\ and \ha\ line widths. The dashed, gray line traces a unity relationship for reference. The data points here are limited to those within 4\arcdeg\ of \lo\ where the \hii\ region dominates emission along the line of sight. Observations with significant multiple-component blending are highlighted in purple; parameters determined from the \hii\ region component fit are less certain for these points.
  \label{fig:widths}}
  \end{center}
\end{figure*}


A grid of the resulting values for $T_i$ and $v$ is
superposed onto Figure~\ref{fig:widths}, a plot of the observed
widths of \ha\ and \sii\ along lines of sight throughout the \hii\
region. Note that as with Figures \ref{fig:icorr} and
\ref{fig:vcorr}, there is general correlation between the widths of
the two lines, suggesting that the same underlying physical
conditions drive the formation of the line profiles of each element
and that the emission arrises from the same region. Although the \hii\ region emission is the dominant component in these spectra within 4\degree of \lo, a few on the edge of the region show noticeable blending with a second component. Examining the ratio of the intensities of the \hii\ region component ($I_{HII}$) to the secondary component ($I_{2}$), we define blended directions as those having $I_{2}/I_{HII} > 0.25$ in either \ha\ or \sii. These directions generally correspond with outliers or points with larger errorbars in Figure~\ref{fig:widths}, which we highlight in purple. 


\begin{figure}[tbp]  
  \plotone{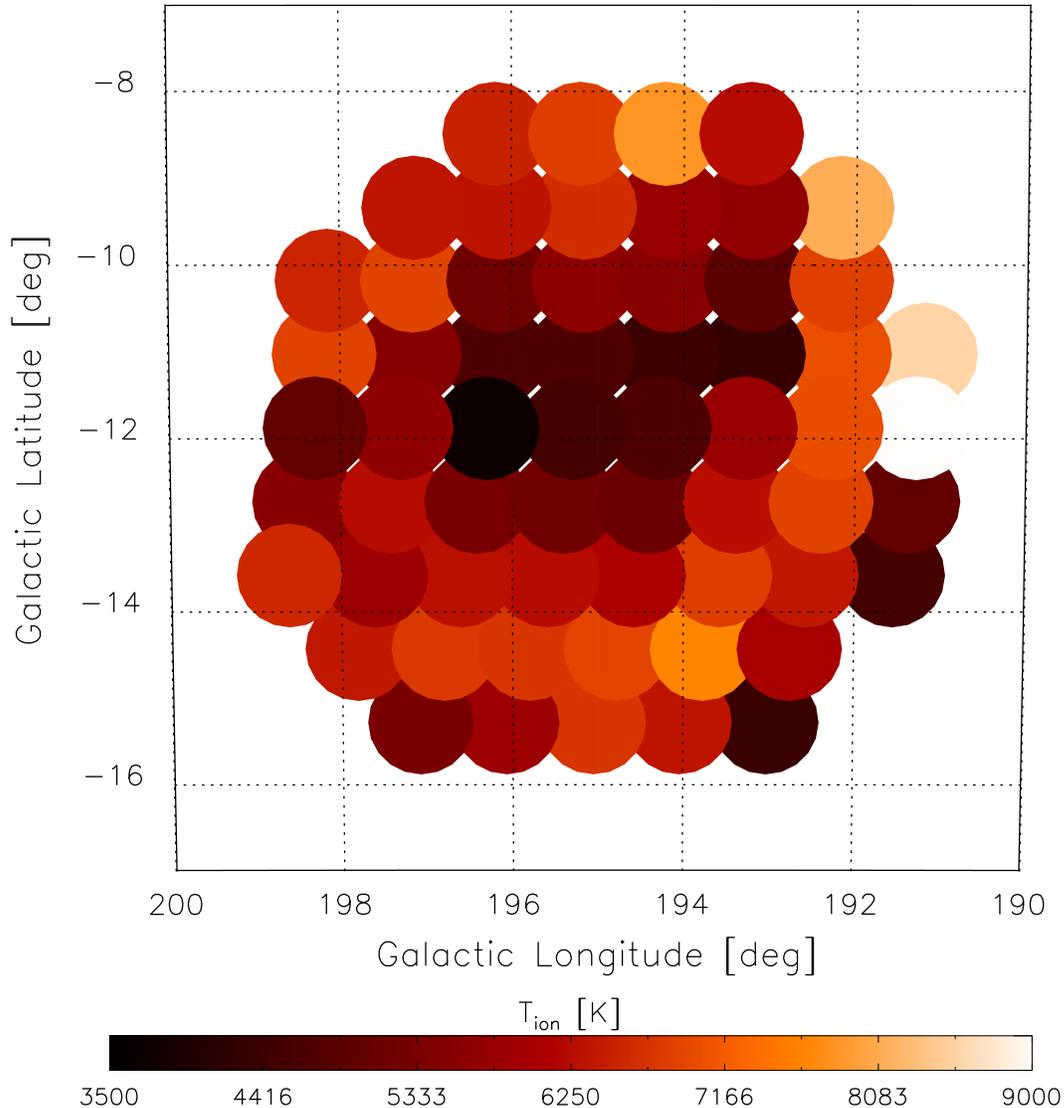}
 \caption{Ion temperature within the \lo\ \hii\ region as derived from the observed \sii\ and \ha\ component widths. Temperatures are depicted as filled circles that show the coverage and sampling of the WHAM 1\arcdeg-beam throughout a region within 4\arcdeg\ of \lo.}
  \label{fig:tmap}
\end{figure}



\begin{figure}[tbp]  
  \plotone{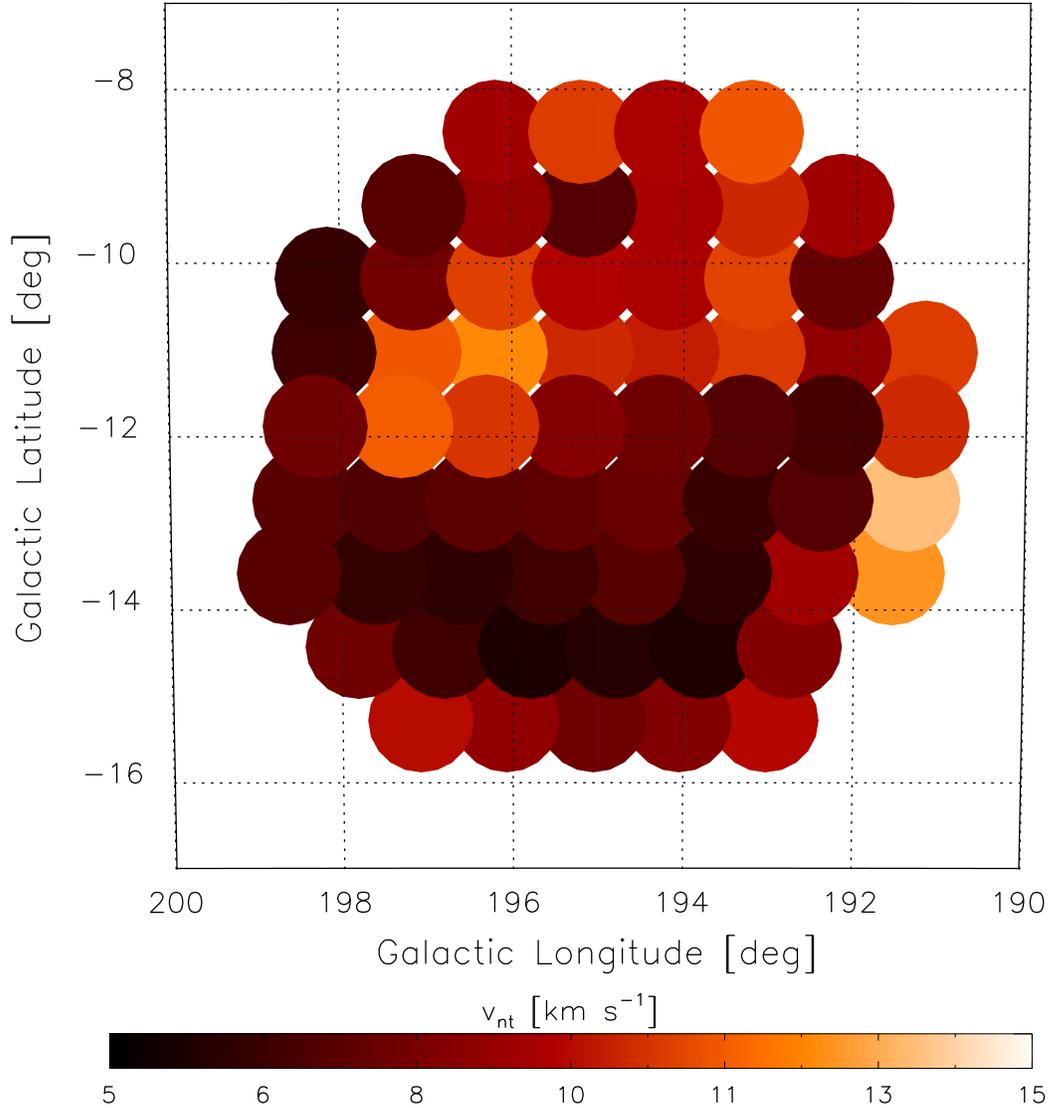}
 \caption{Nonthermal velocity component map of the \lo\ \hii\ region as derived from the observed \sii\ and \ha\ component widths. Velocities are depicted as filled circles that show the coverage and sampling of the WHAM 1\arcdeg-beam throughout a region within 4\arcdeg\ of \lo.}
  \label{fig:vmap}
\end{figure}


We mapped the spatial trends of these temperature and nonthermal
velocity parameters (Figures \ref{fig:tmap} \& \ref{fig:vmap}).
Using this analysis, the ion temperature throughout the nebula
ranges from about 4000~K to 8000~K, with the higher values in the
outer parts of the nebula. Such a trend is not unexpected close to
an ionizing source. Near the source of an H II region the ionization
rate is high and due to their high interaction cross-section,
photons with energy closest to 13.6~eV are used up. Higher energy
photons escape to the edge of the \hii\ region and can heat the
surrounding gas to higher temperatures. \citet{hei00}  derived
similar average temperatures for \hii\ region as a whole by
comparing radio and \ha\ emission.

The nonthermal velocity map (Figure~\ref{fig:vmap}) reveals a
decrease in velocity from about 10 km s$^{-1}$ to 5 km s$^{-1}$ from
the center to the edge of the lower half of the map, perhaps
revealing the expansion of the \hii\ region. In the upper half,
nonthermal velocities are noticeably higher. We note that this area
of increased nonthermal component corresponds to the ``trunks'' of
evaporating material that are more prominent in this portion of the
nebula \citep[e.g., Figure 5 of][]{Ochs15}. With WHAM's limited
spatial resolution in its survey mode, we can only speculate that
ablation and flow from these active regions contributes to a more
complex velocity profile. The nonthermal velocities at the edge of
our studied region appear to increase. This effect could be caused
by emission from the nebula diminishing to an intensity comparable
to emission elsewhere along the line of sight (e.g., from the WIM or
other diffuse portions of the Orion-Eridanus region). When the line profile is no longer dominated by a single component, these derived values become less valid without more complex procedures to isolate the \hii\ region.

The \nii/\ha\ ratio also has the potential to provide an estimate of the electron
temperature throughout the nebula. Combining equations (\ref{eq:em}) and
(\ref{eq:nii}), we have:
\begin{equation}
 {\frac{ I_{6583}}{I_{ H\alpha}}} =1.63\times10^{5} \left({\frac{
 H^{+}}{H}}\right)^{-1}\left({\frac{
 N}{H}}\right) \left({\frac{
 N^{+}}{N}}\right)T_{4}^{0.426}e^{-2.18/T_{4}},
 \label{eq:niiha}
\end{equation}
where $T_{4}$ is electron temperature in units of 10$^{4}$ K and
(N/H) is the gas phase abundance of nitrogen. We use the N/H
gas-phase abundance  of $7.5 \times 10^{-5}$ from \citet{mey97}.
Starting with the simplest assumption that $(H^{+}/H) = 1$ in the \hii\
region, our equation for ratio then becomes:
\begin{equation}
 {\frac{ I_{6583}}{I_{ H\alpha}}} = 12.2\ \left(\frac{N^+}{N}\right)\ T_{4}^{0.426}\ e^{-2.18/T_{4}}
 \label{eq:simp_niiha}
\end{equation}
%


\begin{figure}[tbp] 
  \plotone{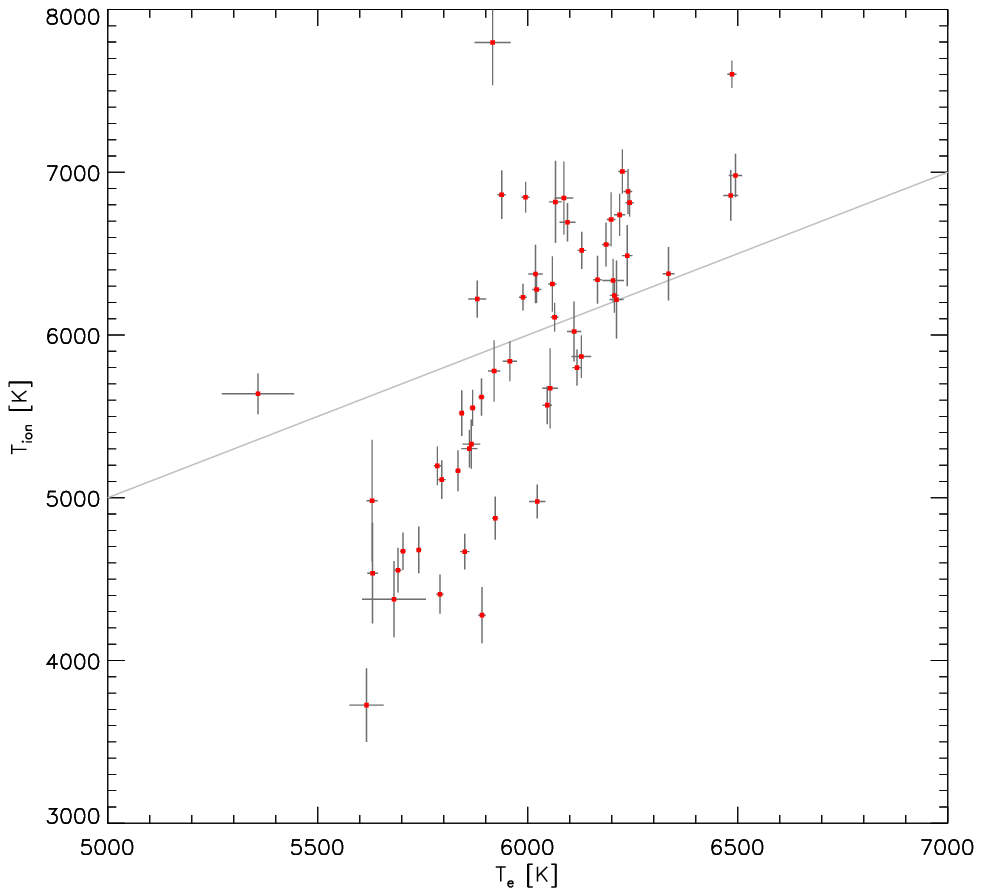}
\caption{Comparison of empirically derived temperature methods. The
ion temperature as derived from \sii\ and \ha\ line widths is
plotted against the electron temperature as derived from the
\nii/\ha\ intensity ratio. For reference, the over-plotted line (\emph{gray})
represents unity. The data points here (\emph{red}) are limited to those within 4\arcdeg\ of \lo\ where the \hii\ region dominates emission along the line of sight; blended spectra are excluded (see text). Error bars (\emph{gray}) represent the propagated $1\sigma$ uncertainty on the temperatures. Note that the uncertainty on $T_e$ from the measured \nii/\ha\ ratio is smaller than the plot symbols for most data points.}
 \label{fig:comparet}
\end{figure}


Using this equation and the \nii/\ha\ line ratio of the \hii\ region
emission components, we can derive an electron temperature at each
location. Unfortunately, $(N^+/N)$ cannot be determined
independently using these observations alone. To provide some basic reference, 
we use $(N^+/N) = 1$ in Figure~\ref{fig:comparet} to compare the
$T_e$ and $T_i$ values from these two different methods. Lowering
$(N^+/N)$ moves points to the right in this plot as higher $T_e$ is
then required to produce the same \nii\ emission. It is clear that
applying a single value of $(N^+/N)$ throughout the nebula will not
lead to better agreement between the two methods. Although the
temperatures do correlate, they span different ranges. For the slope of the data points
between our methods to trend toward unity using the simplified equation~(\ref{eq:simp_niiha}), $(N^+/N)$ would need to decrease with radius. But since the trend in N ionization is important at this point, we should also return to equation~\ref{eq:niiha} and remove the simple assumption $(H^{+}/H) = 1$. Now we see that $(N^+/N)/(H^+/H)$ becomes the factor that must decrease with radius for our methods to agree. This ratio is difficult to isolate with only the lines we present here. The addition of more emission lines maps, such as \oiii, \hei, and \hb, as well as modeling, which will help us examine the changes in ionization state throughout the nebula, are in progress for a future work.

\section{SUMMARY}
\label{Summary}

We have presented multiwavelength observations of \ha, \sii, and
\nii\ toward the \hii\ region surrounding \lo. From these spectra we
have produced velocity-resolved maps. The line-ratio maps that
compare \sii\ and \nii\ to that of \ha\ reveal an increasing trend
that is dependent on radius. While variations in these ratios are
similar to what is observed in the WIM, the ratios in the \hii\
region have significantly lower values. Using the \ha\ intensity
distribution, we have estimated \lc, the Lyman continuum output from
\lo, as well as $n_e$, the average electron density of the \hii\
region. From the widths of \ha\ and \sii, we derive the spatial structure of the ion
temperature (4000 K to 8000 K) and nonthermal velocity (5 \kms\ to 10 \kms) throughout the nebula. We also show a measure of the electron temperature from the relative intensities of
\nii\ and \ha\ and characterize the difference in the two methods of
obtaining temperature. Values generally correlate between the two
methods and reinforce general trends within the nebula. However, the differences (and trend in differences) between the temperatures determined from these methods is larger than explained by measurement error. For such diffuse, locally-ionized regions, the ionization structure changes significantly throughout the nebula that simple assumptions are insufficient to extract reliable temperatures from the intensity ratio method. Relying on observations alone, we find the widths of S and H emission profiles lead to more robust results. Observations of additional emission lines combined with more sophisticated modeling will allow these data to further constrain the ionization structure.

\acknowledgments
 We acknowledge the long-standing support of WHAM by
the U.S. National Science Foundation. The observations and work
presented here were supported by awards AST-0607512 and AST-1108911. This research has made use of the SIMBAD database, operated at CDS, Strasbourg, France.

Facilities: \facility{WHAM}




\end{document}